\documentclass[intlimits,twoside,a4paper]{article}

\usepackage[eqsecnum]{cmpj3}
\usepackage{bm}% bold math

\articletype{A part of the Special collection to the 100th anniversary of the birth of Ihor Yukhnovskii}

\issue{2025}{28}{2}{23604}
\doinumber{10.5488/CMP.28.23604}

\title[Model for smart coating]{Towards construction of microscopic model for smart
coating of a solid surface}
\author[O. Pizio]{O. Pizio\orcid{0000-0001-8333-4652}\thanks{Corresponding author: \email{oapizio@gmail.com}.}}

\address{
Instituto de de Qu\'{i}mica, Universidad Nacional Aut\'{o}noma de M\'{e}xico,
Circuito Exterior, 04510, Ciudad de México, M\'{e}xico}


\date{Received April 17, 2024, in final form June 12, 2024}
\begin{document}

\maketitle

%%%%%%%%%%%%%%%%%
\begin{abstract}

The density functional approach for classical associating fluids is used to
explore the wetting phase diagrams for model systems consisting of water and
graphite-like solid surfaces chemically modified by a small amount of grafted chain molecules.
The water-like fluid model is adopted from the work of Clark et al. [Mol. Phys., \textbf{104}, 3561 (2006)]. It very well describes the bulk water vapor-liquid
coexistence. Each chain molecule consists of tangentially
bonded hard sphere segments. We focus on the investigation of the growth of
water film on such complex substrates and exploration of the
wetting behavior. For grafted monomers, the prewetting phase diagrams
are similar to the diagrams for water on a non-modified solid surface.
However, for grafted trimers and pentamers, a physically much richer behavior
is observed and analyzed.
Trends of the behavior of the wetting temperature and the prewetting critical temperature
on the grafting density and water-segments attraction are discussed in detail.
\printkeywords
%
%
%\pacs 02.70.Ns,61.20.Ja,82.30.Rs,87.15.hp
\end{abstract}

\section{Dedication and history lessons}

The year 2024 has been marked by unfortunate, profoundly sad events for me. On March 26, my teacher
Ihor Rafailovych Yukhnovskii passed away. He kindly, with much patience guided my education 
and growth during first two decades of my scientific career, starting from 1973 at the Department for Statistical Theory of Condensed Systems of the
Institute for Theoretical Physics of the Academy of Sciences of the UkrSSR, 
till my departure to take a new job in Mexico in 1993. 
He tried to convert me into the follower of the ``true faith'' of his collective variables method,
that I recall nowadays as a fascinating endeavour. The present manuscript is dedicated 
to the 100th anniversary of Prof. I. Yukhnovskii birth and as a modest tribute due to his 
important contributions to the theory of soft condensed matter.

This story started at pre-internet times and simultaneously with computer simulations revolution
that influenced the way of thinking of many theoreticians including I. Yukhnovskii.
One of the lines of research initiated by Prof. I. Yukhnovskii was to explore the classical 
ion-molecular fluids under the influence of external field. In fact, this area represents one
of the most important fundamental problems of the microscopic liquid state theory 
that is related to various
applications and technological developments. 

%The first of Yukhnovskii et al. works
The first works of Yukhnovskii and collaborators, published in 1959,
concerned the formulation of the density profiles in terms of the screened potentials
of long-range electrostatic interactions and of the free energy of the model electrolyte of point charges 
as the Mayer-type cluster series~\cite{vladimirov} within the collective variables method. 
The short-range interactions were taken into account by functional differentiation
and contribute as the Boltzmann factors, in, e.g., the virial coefficients of the free energy.
On the other hand, the screening potential of electrostatic interactions was obtained
as the solution of the convolution type integral equation coming from the approximation
for the transition Jacobian from the Cartesian coordinates to the collective variables.

This pioneering work appeared at the same period of time as the Percus and Yevick
integral  equation for a hard sphere fluid model~\cite{percus} and much earlier
than the statistical mechanical procedure  proposed by Henderson, Abraham 
and Barker~\cite{doug1} to describe simple fluids in contact with hard solid wall. 
This kind of approach was extended to systems with Coulomb inter-particle interaction
besides hard sphere repulsive forces, by Blum et al.~\cite{lesser1,lesser2} almost immediately
at the level of the mean spherical approximation.

By contrast, Yukhnovskii and his younger co-workers persuaded the development of
the collective variable method to inhomogeneous fluids with long-range 
electrostatic interactions considering electrolyte solution type models
and assuming the presence of a sharp uncharged boundary and the dielectric discontinuity 
in the system~\cite{kuryliak1,kuryliak2,sovjak1,sovjak2}.

At the early seventies of the last century (around 1970--1972), I. Yukhnovskii 
reconsidered the developments from the integral equations method for simple fluids 
with dominating short-range repulsive interaction, principally under the influence
of his long-term co-worker Prof. M. Holovko. 
As a result, the reference system reformulation of the transition Jacobian to collective
variables was developed by Yukhnovskii. Within this methodological procedure, the
screening potentials follow from the Ornstein-Zernike type integral equations both
for homogeneous and non-homogeneous ion-molecular fluids. 
Next, the expressions for the free energy and pair distribution functions are
obtained straightforwardly (as in the original version of the collective variables method).
These series were termed as the optimized cluster expansions.
One example of the application of this procedure for the ion-dipole model electrolyte
by using the screening potentials from the solution of the mean spherical 
approximation is given in~\cite{blotsky}. 

A comprehensive account of the versions of the collective variables method together with
the integral equations theories for various model fluids with the emphasis on ion-molecular
mixtures is given in the extraordinarily important monograph by 
Yukhnovskii and Holovko~\cite{yukhnovskyi1}. This book provides a profound
analysis of the relations between different methods and regarding the limits of their applicability.
Moreover, the book outlined necessary developments that in fact inspired various
posterior works. A brief account of the results concerning  the screening potential for inhomogeneous
ion-molecular model fluids of point-like particles is also available in~\cite{golovko,holovko}.

In 1991 Ukraine regained its independence. I.~R.~Yukhnovskii, as a prominent scientist and
one of intellectual leaders of the nation, interrupted his scientific activities
becoming involved in the top governmental bodies and the parliament of the new state.
The ensuing efforts in the construction of the theory of inhomogeneous fluids, mixtures and
electrolyte solutions were undertaken by the academic stuff grown in the Department for Theory of Solutions of the ICMP (%Institute of Condensed Matter Theory of the National Academy of Sciences of Ukraine
Institute for Condensed Matter Physics of the National Academy of Sciences of Ukraine) guided by Prof. M. Holovko at that time.

Principal efforts in the area of research were undertaken in the extensions of the available methods
to deal with strong inter-particle attraction or, in other words, to take into account
the chemical association of species. Specifically, theoretical foundations of the approach
by Wertheim in terms of multi-density thermodynamic perturbation theory and integral equations of the
Ornstein-Zernike type for differently associated species in the bulk, homogeneous fluids
served for this purpose. To be brief and closer to the objective of the present manuscript,
I would like to refer to some first studies concerning inhomogeneous associating
fluids~\cite{vakarin1,vakarin2,stefan1}. In contrast to these, early studies 
using integral equations methodology, the ensuing efforts from this laboratory were 
performed within density functional approaches for associating fluids~\cite{chapter}.
It is worth mentioning that general aspects of the relation between collective variables and
density functional approaches were elucidated some time ago~\cite{mryglod}.

Concerning our applications of this technique to systems with non-electrostatic 
inter-particle interactions, we explored the two-site and four-site fluid models 
with site-site chemical association, besides Lennard-Jones inter-particle attraction,
in contact with various solid surfaces and in pores of 
different geometry~\cite{huerta,malo,malo2,anpat1}.
Our interest to revisit all these findings and explore novel systems, in part, had arisen 
due to a successful parametrization of the four-site water model with square well 
non-associative attraction, in the laboratory of G. Jackson~\cite{clark}. 
The model and method permitted to obtain novel insights into adsorption and phase behavior
of water in nanoscopic slit-like pores~\cite{trejos1,trejos2}.
More recently, we considered the effects of fluid-solid interaction strength
on the wetting of graphite-like solids by water~\cite{opss1}.
In that work, the temperature dependence of the contact angle 
for water on different substrates was obtained from the Young equation besides the exploration
of adsorption isotherms. Trends of behavior of the contact angle  permitted to make estimates 
for the wetting temperature of water as well.
With this knowledge, we have attempted to study the phase behavior, adsorption, 
wetting temperature and contact angle for several systems that involve more 
complex substrates. Namely, we considered the behavior of water in contact with
graphite-like solid surfaces that are chemically modified by grafting of chain molecules 
and their mixtures~\cite{opss2,opss3}. As a by-product, the contact angle of water model
on solids with grafted monomer species was explored as well~\cite{dabrowska}. 
Unfortunately, on June 24 of 2024, my very close friend, Stefan Sokolowski, 
with whom we permanently worked
together during three decades, since 1994, suddenly passed away~\cite{obss}.  
All theoretical background used
to elaborate the present manuscript is the product of our common efforts.
The principal scientific objective of this manuscript and formal introduction 
are given in the next section.

\section{Introduction}

Chemically modified solid substrates are common objects in contemporary
surface science. They are intentionally designed to produce surfaces with
desired specific properties or functionalities. Grafting chain particles 
(oligomers and/or polymers) is one of the powerful methods
to yield functionalized surfaces that have an enormous area of applications.
%control and alter the wettability of solid surfaces.
One kind of systems resulting from such a procedure is termed as polymer 
brushes~\cite{minko}. As an opposite extreme, a rather small molecule
can be used for surface functionalization~\cite{verena,guti,obeso}
with quite different purposes.
One example of frequently
used modifiers of silica-based surfaces is 3-Aminopropyltriethoxysilane 
(APTES) molecule~\cite{guti,obeso}. It can be
viewed (in coarse grained representation) as a molecule composed of four segments. 
APTES can covalently attached to the silica
surface through the formation of siloxane bonds, whereas its amine group
extends away from the substrate. This type of modified solid surfaces, in general terms, 
is used for selective adsorption of undesirable species from aqueous 
media. However, adsorption of the dominant component, i.e., of water solvent, 
upon the changes of the conditions of the experiment or process, should be known 
prior to the study of aqueous solutions.  

Trends of behavior of adsorption can be interpreted in terms of wetting.
Majority of experimental and theoretical studies in the literature 
concern the wetting of water by smooth homogeneous solid substrates. 
The influence of substrate heterogeneity, at 
macro or mesoscopic scale, on the adsorption and wetting has been considered 
in several publications as well. From the previous analyses, it is 
expected that the wetting properties of modified surfaces due to grafted molecules 
are different, in comparison with bare (or non-modified solids) \cite{j1,j2,j3}.

Usually, the wetting transition, between non-wetting and wetting behavior,
is elucidated by two methods.
One of them is based on the measurement of the contact angle,
$\theta$. The wetting transition takes place when $\theta$ drops from a
non-zero value to zero. The characteristic temperature at which such a change 
occurs is the wetting temperature. 
On the other hand, another experimental method to determine the wetting
temperature relies on the evaluation of the adsorption isotherms 
for gas densities up to the density of the bulk saturated vapor
at different temperatures. Then, one can
recover the line of the first-order prewetting
transition that starts at the wetting temperature and ends up
at the prewetting critical temperature for, e.g., strongly attractive substrates.
It is important to mention that the method based on adsorption isotherms
permitted to obtain experimental wetting temperature and the
prewetting transition temperature for $^4$He on cesium for the first
time \cite{w1}. Both these methods formally can be implemented within the relevant 
theoretical procedures. One can  evaluate the quality of modelling the system 
with respect to the indicators of its interfacial behavior.

%%%%%%%%%%%%%%%%%%%%%%%%%%%%%%%%%%%%%%%
In fact this task is not straightforward. The wetting behavior follows from
an interplay between fluid-fluid and fluid-solid interaction, besides
the interactions involving segments of grafted molecules with
fluid species and with solid surface. Finally, the interaction between
segments of different grafted molecules matter as well.
Consequently, certain assumptions or simplifications are inevitable within
theoretical modelling.
If the grafted molecules are long chains, mimicking polymer species, and if
one restricts to the brush regime, the interactions of fluid species 
and of exposed segments with solid substrate can be neglected.
On the other hand, if the grafted species are short,
one should find arguments to choose the values of parameters for
the entire set of interactions.

Theoretical approaches to the study of adsorption of fluids on solids
modified with grafted chain molecules are usually based
on different versions of density functional (DF) methods~\cite{df2,df3,df4,df5,df6,df7}.
Our approach, as in the previous works from this laboratory, is based 
on the theory of adsorption of a mixture of chain molecules,
developed by Yu and Wu \cite{wu1}.
The theory is
successful in describing the wetting and layering transition \cite{wet1,wet2},
as well as phase transitions in fluids confined to nanoscopic slit-like pores
with modified walls~\cite{n7,n8}.

In our recent work \cite{opss1}
we  investigated a prewetting transition for water adsorbed on
graphite-like substrates using the density functional approach
for classical associating fluids. Water-water interactions were modelled
according to the SAFT approach \cite{saft}. The fluid-solid interaction, however,
is assumed to have a 10-4-3 Lennard-Jones type potential form \cite{steele}.
The interaction energy of the latter potential
is considered to be a
free parameter while the wetting and the prewetting critical temperatures as well as the prewetting
phase diagrams are evaluated for different fluid-solid attractions.
Theoretical findings are in agreement with the experimental
results \cite{es1} as well as with data from computer simulations \cite{es2,es3,es4,es5}.
In spite of less sophisticated modelling of the system within the density functional
approaches in comparison with computer simulations,
theoretical calculations provide a rather comprehensive description of the wetting
phenomena. Theoretical calculations are computationally less expensive 
and faster than computer simulations performed with the same goal.
Therefore, the density functional approach provides a convenient tool
to study the adsorption and phase behavior of associating fluids on the
solids modified with tethered chain molecules.

The principal aim of this paper is to study the wetting behavior of
water-like species on graphite-like surfaces modified by 
short tethered chain  molecules.
Then, the  phase diagrams can be compared with their counterparts for
non-modified surfaces \cite{opss1}. 
The present work is
a continuation of our previous studies of adsorption and wetting of
surfaces with tethered chains \cite{wet1,wet2}. 
We restrict ourselves to the study of the models of grafted species (monomers, trimers and
pentamers) built of hard-sphere segments. However, the segments
of grafted species may attract water molecules.
The principal issues of the study lie on the influence of the grafting density 
and the number of segments of the grafted molecules on the prewetting transition.
A comparison of adsorption trends of water and the wetting behavior 
for grafted and ungrafted models is performed.

\section{Model}

We study a model for adsorption of water  on graphite-type solid
modified with end-grafted chain molecules. 
The fluid particles and grafted molecules are considered as a mixture 
of species put in contact with a
wall. The grafted molecules are the chains built of $M$ tangentially bonded spherical 
hard segments of diameter $\sigma_\text{s}$.
The connectivity of segments belonging to a given molecule is
provided by imposing the bonding potential, $V_B$~\cite{wu1},
\begin{equation}
 \exp[-\beta V_B(\mathbf{R}]=
 \prod_{i=1}^{M-1}\delta(|\mathbf{r}_{i+1}-\mathbf{r}_i|-\sigma_\text{s})/4\piup\sigma_\text{s}^2,
\end{equation}
where $\mathbf{R}\equiv (\mathbf{r}_1,\mathbf{r}_2,\dots \mathbf{r}_M)$ is the vector
specifying the positions of segments, $\delta$ is the Dirac function, and $\beta=1/k_\text{B}T$.

The first segment of each chain molecule is irreversibly bonded to the
substrate by the potential, see, e.g., \cite{n7,n8,n1,n3},
\begin{equation}
 \exp[-\beta v_{\text{s}_1} (z)] = {\cal C}\delta(z-\sigma_\text{s}/2),
\end{equation}
where $z$ is a distance from the surface and ${\cal C}$ is a constant.
Thus, the model implies that the surface-binding segments are at
a fixed distance from the surface along $z$-axis although they can slide 
within the $xy$-plane.  
The solid surface is an impenetrable hard wall for all other segments
$i=2,3,\dots,M$,
\begin{equation}
 v_{\text{s}_{i}}(z)=\left\{
 \begin{array}{cc}
  \infty, & z\leqslant \sigma_\text{s}/2, \\
  0,   &     z>\sigma_\text{s}/2.
 \end{array}
\right.
\end{equation}

The interactions between water
molecules are described by the model from 
\cite{jackson1,jackson2} parameterized by Clark et al. \cite{clark}.
As the rationale for the choice of the interaction potential was given
in \cite{clark}, we would like to recall solely a few issues.
Each fluid molecule possesses four associative sites denoted as A, B, C, and D,
inscribed into a spherical core.
The set of all the sites is denoted by $\Gamma$.
The pair interlocutor potential between molecules 1 and 2  depends
on the center-to-center distance and orientations,
\begin{equation}
 u(r_{12}) = u_{\text{ff}}(r_{12}) + \sum_{\alpha \in \Gamma} \sum_{\beta \in \Gamma} 
u_{\alpha\beta}(\mathbf{r}_{\alpha\beta}),
\end{equation}
where
$\mathbf{r}_{\alpha\beta}=\mathbf{r}_{12}+\mathbf{d}_{\alpha}(\omega_1)-\mathbf{d}_{\beta}(\omega_2)$
is the vector connecting site $\alpha$ on molecule 1 with site $\beta$ on molecule 2,
 $r_{12}=|\mathbf{r}_{12}|$ is the distance between centers of molecules 1 and 2,
$\mathbf{\omega}_i$ is the orientation of the molecule $i$, $\mathbf{d}_{\alpha}$
is the vector from the molecular center to the site $\alpha$,
see, e.g., figure~1 of~\cite{jackson1} for better visualization.
Each of the off-center attraction sites is located at a
distance $d_{\text{s}}$ from the particles' center, $d_\text{s} = |\mathbf{d}_{\alpha}|$
($\alpha = \text{A, B, C, D}$).

Only the site-site association  AC,  BC, AD, and BD is allowed and all
site-site association energies are assumed equal. The interaction between sites
is given as
\begin{equation}
\label{eq:asw}
u_{\alpha\beta}(\mathbf{r}_{\alpha\beta})=
\left\{
\begin{array}{ll}
-\varepsilon_{\text{as}}, & {\rm  if \ \ } 0< |\mathbf{r}_{\alpha\beta}| \leqslant r_\text{c} ,\\
 0,  & {\rm  if \ \ }   |\mathbf{r}_{\alpha\beta}|   > r_\text{c} ,
\end{array} \right.
\end{equation}
where $\varepsilon_{\text{as}}$ is the depth of the association energy well and
$r_\text{c}$ is the cut-off of the associative interaction.

The non-associative part of the pair potential, $u_{\text{ff}}(r)$, is given as
\begin{equation}
u_{\text{ff}}(r) = u_{\text{hs},\text{ff}}(r) +  u_{\text{att},\text{ff}}(r),
 \label{eq:sw}
\end{equation}
where $u_{\text{hs},\text{ff}}(r)$ and $u_{\text{att},\text{ff}}(r)$ are the
hard-sphere (hs) and attractive (att) pair interaction potential, respectively.
The hs term is,
\begin{equation}
u_{\text{hs},\text{ff}}(r) =
\left\{
\begin{array}{ll}
\infty, &  {\rm if \ \ } r < \sigma,\\
0,   &  {\rm if \ \ } r \geqslant \sigma,
\end{array}
\right.
\end{equation}
where $\sigma$ is the hs diameter.
The attractive interaction is described by the square-well  potential,
\begin{equation}\label{uSW}
 u_{\text{att},\text{ff}}(r)=
\left\{
\begin{array}{ll}
0, & {\rm if \ \ } r < \sigma,\\
-\varepsilon, & {\rm if \ \ } \sigma \leqslant r < \lambda \sigma,  \\
0, & {\rm if \ \ } r \geqslant \lambda \sigma,
\end{array}
\right.
\end{equation}
where $\varepsilon$ and $\lambda$ are the depth and the range of the
non-associative attraction potential, respectively.

According to the model design \cite{clark}, four sets of parameters
of the potentials from equation~(\ref{eq:asw}) and equation~(\ref{uSW}) were proposed.
They are denoted as W1, W2, W3, and W4 models.
The W1 model slightly better reproduces the bulk phase diagram and
the temperature dependence of the surface tension of water
than the W2, W3 and W4 models, see figure~2 of \cite{wet2} and
data resorted in \cite{clark,opss1} for these properties.
%}
The parameters of the W1 model are given here in table \ref{tab:1} for the 
convenience of the reader.

\begin{table}[h]
  \centering
   \caption{
  The parameters of the W1 water-water model potential
   from~\cite{clark}.
   }
   \vspace{0.3cm}
   \begin{tabular}{ccccccc
   }
  \hline \hline
 Model &  $\sigma  $ (nm)& $(\varepsilon/k)$ (K) & $\lambda$&  $r_\text{c} $ (nm)&
     \vspace{0.1cm} $(\varepsilon_{\text{as}}/k)$ (K) & $d_s/\sigma$  \\[0.5ex]
     \hline
W1 & 0.303420 & 250.000 & 1.78890 & 0.210822 & 1400.00 & 0.25 \\
\hline \hline
\end{tabular}
\label{tab:1}
\end{table}

The interaction of water molecules  with carbonaceous solids
is described by the potential developed by Steele \cite{steele}.
It reads,
\begin{equation}\label{steele}
 v_{\text{sf}}(z) =2\piup\rho_\text{g}\varepsilon_{\text{sf}}\sigma^{2}_{\text{sf}}\Delta
 \left[ \frac{2}{5} \left( \frac{ \sigma_{\text{sf}}}{z}\right)^{10}
 - \left( \frac{ \sigma_{\text{sf}}}{z}\right)^{4} \right.
% \nonumber \\ &
\left. - \frac{\sigma_{\text{sf}}^4 }
{3 \Delta (z+0.61 \Delta)^3 }
 \right],
\end{equation}
where $\varepsilon_{\text{sf}}$, $\sigma_{\text{sf}}$ are
the energy and the distance parameters, respectively; $\Delta$ is the interlayer spacing
of the graphite planes, $\Delta= 0.335$~nm and $\rho_\text{g}$ is the density of graphite,
$\rho_\text{g}= 114~ {\rm nm}^{-3}$. To calculate the value of  $\sigma_{\text{sf}}$
we applied Lorentz additivity rule, $\sigma_{\text{sf}} = (\sigma_\text{g} +\sigma)/2$,
where $\sigma_\text{g}=0.34$~nm is the diameter of carbon atoms in graphite.
Some aspects of the applicability of equation~(\ref{steele}) are discussed
in~\cite{steele,es5}. On the other hand, recently we
showed that this potential can be successfully applied to describe the temperature dependence
of the contact angle of water on graphite-like solids \cite{opss1}.

The interaction of water particles 
with each segment of grafted molecules is assumed
in the form, $u_{\text{fc}}(r) = u_{\text{hs},\text{fc}}(r) +  u_{\text{att},\text{fc}}(r)$. It is
the same as for non-associative water-water interaction.
In the hard sphere term, we choose $\sigma_{\text{fc}} = \sigma_\text{s} = \sigma$
for simplicity. 
The attractive contribution is taken in the square-well form, similarly
to equation~(\ref{uSW}), although with the parameters $\lambda_{\text{fc}}$ and $\varepsilon_{\text{fc}}$.
A reasonable choice of $\lambda_{\text{fc}}$ and $\varepsilon_{\text{fc}}$ is not straightforward.
It should be based on any kind of experimental information that we lack at present.
Finally, we assumed that the interaction between segments of different grafted
chains is of hard sphere type with the parameter $\sigma_\text{s}$.

The grafting density of chain molecules is $r_\text{c} = N_\text{s}/A$,  $N_\text{s}$ is the number of
grafted molecules and $A$ is the surface area. The
system under study is considered in equilibrium with the reservoir containing water
molecules only. The bulk density and the chemical potential of water
are denoted as $\rho_\text{b}$ and $\mu$, respectively.

\section{Theory: density functional approach}

We use the density functional approach based on the theory
originally developed by Yu and Wu \cite{wu1}. This approach has
been already used to describe systems involving grafted
layers \cite{n7,n8,n1,n3,opss1}.
The free-energy functional is constructed in the perturbation manner, i.e., it is
the sum of the contributions arising from
different interactions in the system, $F = F_{\text{id}} + F_{\text{hs}} + F_{\text{c}} + F_{\text{as}}+ F_{\text{att}}$.
The ideal part of the free
energy, $F_{\text{id}}$, is known exactly \cite{wu1}. The excess free energy due to
hard-sphere interactions, $F_{\text{hs}}$, follows from the White Bear version of the Fundamental
Measure Theory \cite{wu2}.  The chain connectivity
contribution, $F_\text{c}$, as well as the contribution corresponding to the
associative water-water interaction, $F_{\text{as}}$, result from the first-order perturbation theory
of Wertheim \cite{jackson1,jackson2,wu2}.

Each of the above mentioned contributions is defined as a functional of the local density of
fluid molecules, $\rho(\mathbf{r})$, and the densities of particular
segments $\rho_{\text{s}_{i}}(\mathbf{r})$.
These contributions were described in detail
in \cite{wu1,wu2}.
In addition, it is convenient to define the total segment density profile,
\begin{equation}
\rho_{\text s}(\mathbf{r})=\sum_{i=1}^M \rho_{\text{s}_{i}}(\mathbf{r}).
\end{equation}

The  attractive free energy term
results from  the mean-field approximation,
\begin{equation}
 F_{\text{att}}=\frac{1}{2}\int \rd\mathbf{r}_1 \rd\mathbf{r}_2
 \rho(\mathbf{r}_1)  \rho(\mathbf{r}_2)u_{\text{att},\text{ff}}(|\mathbf{r}_1- \mathbf{r}_2|)\\
+\sum_{i=1}^M\int \rd\mathbf{r}_1 \rd\mathbf{r}_2\rho(\mathbf{r}_1) \rho_{\text{s}_{i}}(\mathbf{r}_2)u_{\text{att},\text{fc}}(|\mathbf{r}_1- \mathbf{r}_2|),
\end{equation}
where, $u_{\text{att},\text{ff}}(r)$ coincides with $u_{\text{att},\text{fc}}(r)$ introduced in the previous subsection.
With the free energy constructed, the essence of the calculations
is in the minimization
of the thermodynamic potential. The
grafted density is fixed by the constraint,
\begin{equation}
\int \rd z \rho_\text{s}(z) = R_\text{c}.
\label{eq:con}
\end{equation}

In the above equation we have taken into account that for one-dimensional external field as in
equation~(\ref{steele}), the local densities of segments and fluid molecules
depend on the coordinate $z$ only. If the constraint (\ref{eq:con}) is taken
into account, the thermodynamic potential is given as,
\begin{equation}
 {\cal Y} = F[\rho_\text{s}(z),\rho(z)] +\sum_{i=1}^M\int \rd z\rho_{\text{s}_{i}}(z)v_{\text{s}_{i}}(z)+
 \int \rd z \rho(z)[v_{\text{sf}}(z)-\mu].
\end{equation}

The Euler-Lagrange
equations (cf. \cite{opss1,wu2}) that follow from the
minimization of the thermodynamic potential,
\begin{eqnarray}
  \frac{\delta {\cal Y}}{\delta \rho(z)} =0, \\
 \frac{\delta {\cal Y}}{\delta \rho_{\text{s}_{i}}(z)} =0, 
\end{eqnarray}
are solved numerically using Picard iteration algorithm.
The solution yields the density profiles $\rho(z)$, and $\rho_{\text{s}_{i}}(z)$ for $i=1,2,\dots,M$ and
the total profile of grafted species $\rho_\text{s}(z)$ as a result.  

The density profile $\rho(z)$ determines the excess adsorption isotherm,
\begin{equation}
 \Gamma_{\text{ex}}=\int \rd z [\rho(z)-\rho_\text{b}].
\end{equation}
The behavior of the grafted layer upon changing the external conditions can be described by 
using the first moment of the density profile $\rho_\text{s}(z)$.
In the case of long grafted species in the brush regime, it is 
interpreted as a brush height, $\langle h_{\text{c}}\rangle$. It can
be defined as \cite{he1,he2},
\begin{equation}
 \langle h_{\text{c}}\rangle=2\int \rd z z \rho_\text{s}(z)\big/\int \rd z \rho_\text{s}(z).
\end{equation}

Similarly to our recent study \cite{opss1}, the
localization of the equilibrium transition along each
isotherm relies on the calculations of the excess
 thermodynamic potential,
 ${\cal Y}-\Omega_\text{b}$, where $\Omega_\text{b}$
denotes the grand canonical thermodynamic potential of the bulk fluid,
along the increasing and  decreasing chemical
potential paths, starting the iterations from
the density obtained at a previous value of the chemical potential.

Except otherwise stated, all the quantities are given
in reduced units. The parameters characterizing water species,
$\sigma$ and $\varepsilon$, are chosen as the length and
energy units.
The reduced temperature then is
$T^*=k_\text{B}T/\varepsilon$,  the reduced distance is
$z^*=z/\sigma$, $R^*_\text{c}=R_\text{c} \sigma^2$, $\langle h^*_{\text{ch}}\rangle=\langle h_{\text{ch}}\rangle/\sigma$,
the reduced adsorption and reduced local density are
$\Gamma_{\text{ex}}^*=\Gamma_{\text{ex}}\sigma^2$, $\rho^*(z)=\rho(z)\sigma^3$
(similarly, $\rho_\text{s}^*(z)=\rho_\text{s}\sigma^3$).
In addition, $\varepsilon^*_{\text{gs}} = 2\piup\rho_\text{g}\varepsilon_{\text{sf}}\sigma^{2}_{\text{sf}}\Delta/\varepsilon$,
and $\varepsilon^*_{\text{fc}}=\varepsilon_{\text{fc}}/\varepsilon$.

For a given model of interaction between water molecules,
the system is characterized by  parameters. They are:
the diameter of spherical segments, the number of segments $M$, the grafting
density, and the parameters of the water-solid potential from equation~(\ref{steele}).
Principally, our calculations focused on the evaluation of the wetting properties, concern the changes
of grafting density, $R^*_\text{c}$, for molecules composed of a different number of segments, $M$, 
and the interaction energy between water particles and segments of grafted molecules,
$\varepsilon^*_{\text{fc}}$.

\subsection{Results}

Intuitively,  one could expect that the influence
of a small amount of grafted molecules on the prewetting phase behavior
should be similar to the effect of decreasing the value of water-solid
interaction strength, $\varepsilon_{\text{gs}}^*$.  We would like to explore this issue
more in detail. 

\begin{figure}[h!]
\begin{center}
\includegraphics[width=6.5cm,clip]{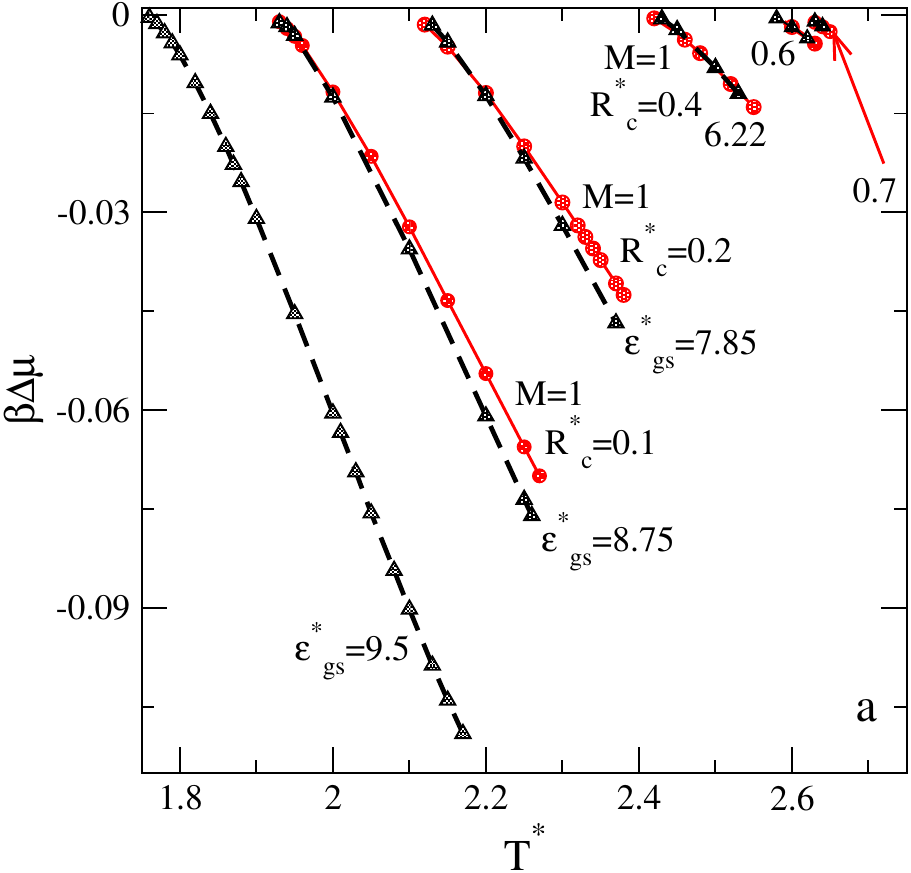}
\includegraphics[width=6.5cm,clip]{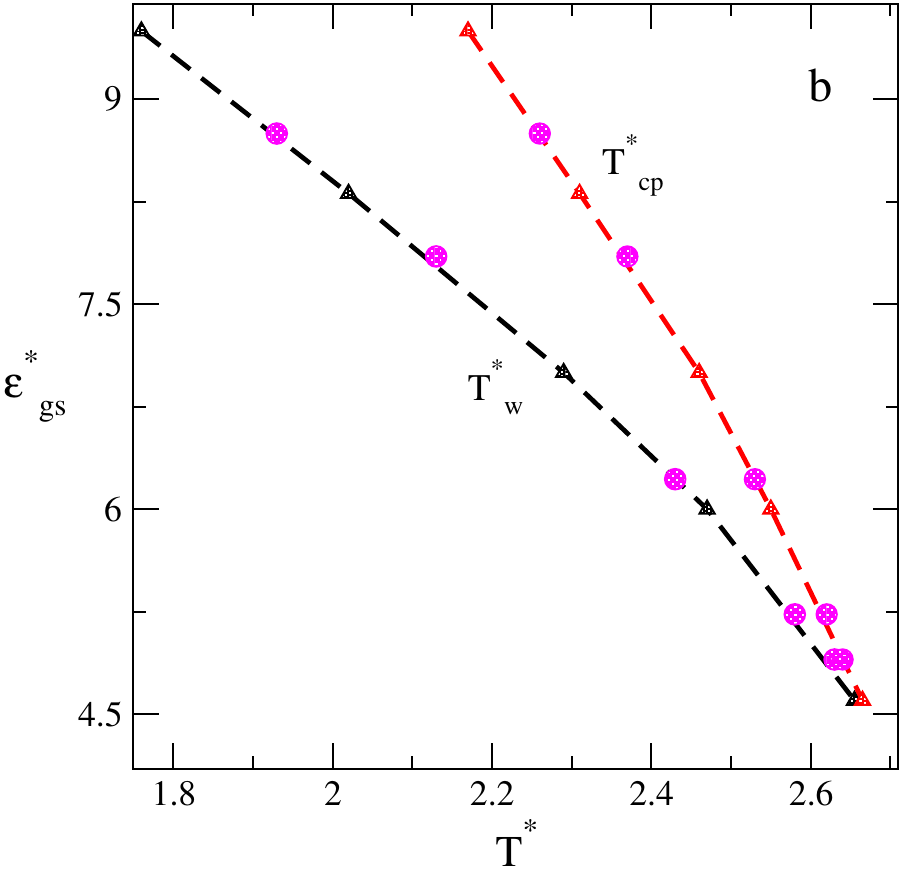}
\includegraphics[width=6.5cm,clip]{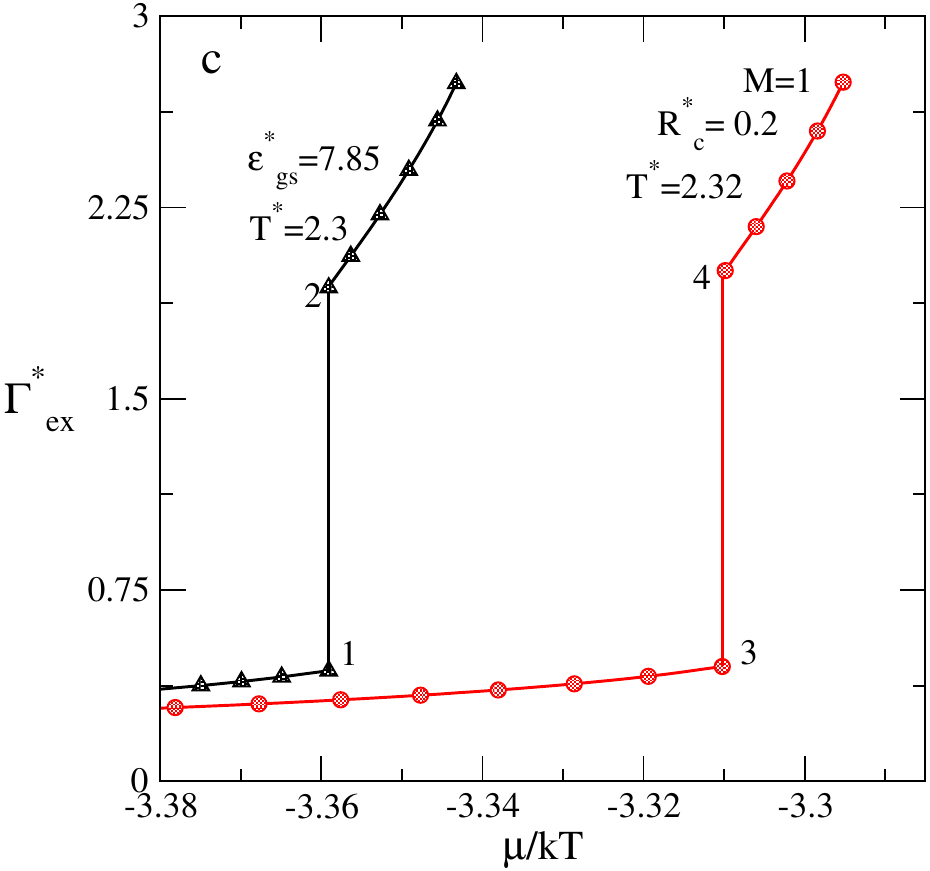}
\includegraphics[width=6.5cm,clip]{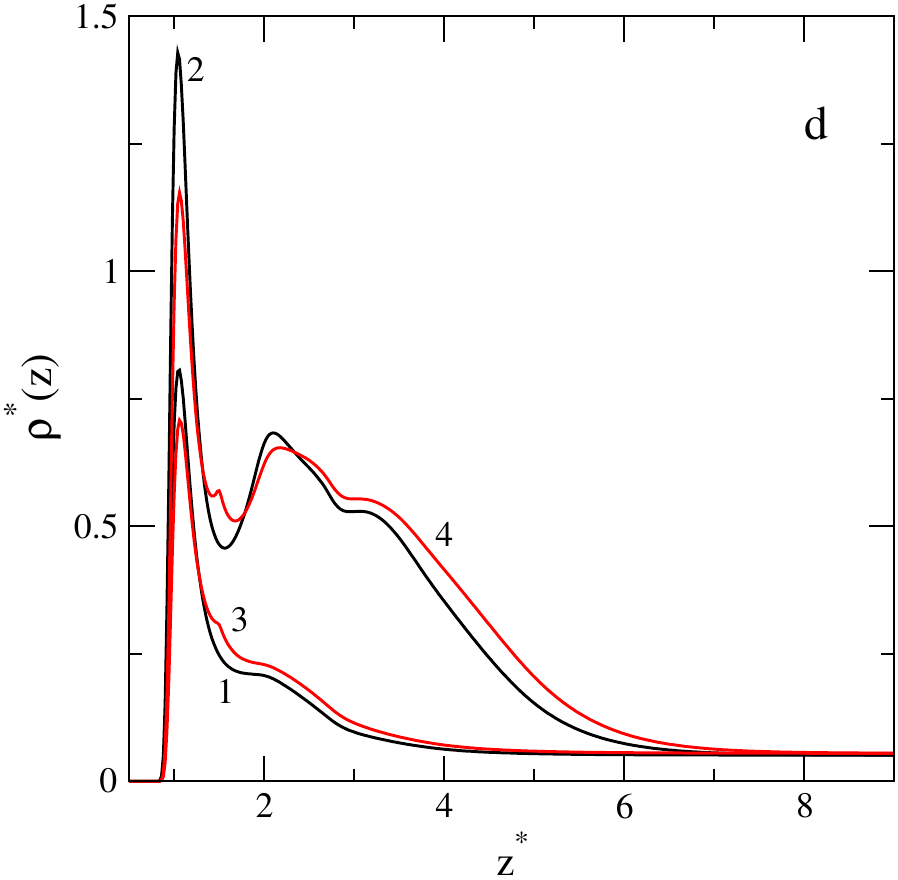}
\end{center}
\caption{(Colour online) Panel a: 
The prewetting phase diagrams for
water on graphite-like solid with different adsorbing strength
($\varepsilon_{\text{gs}}^*=9.5$, 8.75, 7.85, 6.22, 5.23 and 4.9, 
from the left to right, plotted as dashed lines
with triangles). The last two numbers are omitted to make the
figure less loaded.
The  diagrams for the model with grafted monomers
at $R^*_\text{c} =$0.1, 0.2, 0.4, 0.6 and 0.7 are plotted as red solid lines
with circles.
Panel b: The characteristic temperatures, $T^*_\text{w}$ and $T^*_{\text{cp}}$
of the prewetting phase diagrams for two sets of models in
$\varepsilon_{\text{gs}}^*$ -- $T^*$ plane. Dashed lines with triangles
correspond to graphite-like solid with different adsorbing strength.
Magenta circles indicate $T^*_\text{w}$ and $T^*_{\text{cp}}$ for the models with
grafted monomers as in panel a.
Panel c: Fragments of adsorption isotherms of two models illustrating the
prewetting phase transition.
Panel d: The density profiles of water species just before and after the
prewetting transition at states numbered in panel c.
}
\label{fig1}
\end{figure}

To begin with, let us consider a set of results concerning the
system of monomers, $M=1$, grafted on a strongly adsorbing graphite-like solid
with $\varepsilon_{\text{gs}}^*=9.5$. Besides, we assume that the interaction
between grafted monomers and water particles is of hard sphere 
type, i.e., $\varepsilon_{\text{fc}}^*=0$.
If the grafted density changes from zero (bare substrate with a vanishing amount of grafted species)
up to $R^*_\text{c} = 0.7$, the wetting temperature, $T^*_\text{w}$, changes from $T^*=1.76$ (bare substrate) 
to a high value, $T^* \approx 2.63$, quite close to the critical temperature
of bulk water ($T^*_\text{c} \approx 2.718$). This is illustrated in the chemical potential-temperature
plane, $\beta (\mu - \mu_0)$ ($\mu_0$ is the chemical potential at bulk coexistence), figure~\ref{fig1}a.
The lines of the first-order prewetting transition, plotted in red
with circles in each case, start from $T^*_\text{w}$ and end up at a certain critical prewetting 
temperature $T^*_{\text{cp}}$, figure~\ref{fig1}a. At $T^* < T^*_\text{w}$, water does not wet the surface.
On the other hand, at $T^*_{\text{cp}} < T^* < T^*_\text{c}$, the adsorption isotherms are smooth. 
They tend to infinity upon approaching the bulk coexistence.
Next, I performed calculations of the prewetting phase diagrams of non-grafted solid
surfaces searching for the values of $\varepsilon_{\text{gs}}^*$ that yield a similar value of the
wetting temperature as for grafted systems. These results are plotted as black lines
decorated by triangles.
It can be seen that the lines describing two sets of phase 
diagrams nearly coincide, figure~\ref{fig1}a. 
A small difference is observed, however, close to the critical temperature of prewetting. 
Still, if one constructs the curves for two
characteristic temperatures,  $T^*_\text{w}$ and $T^*_{\text{cp}}$, in the $\varepsilon_{\text{gs}}^*$ -- $T^*$
plane, the magenta circles (corresponding to grafted systems with different $R^*_\text{c}$) fit well
on the curves describing the substrates of a different adsorbing strength without grafted 
monomers, figure~\ref{fig1}b. From these results one can get an idea about the changes of wettability of
the grafted systems using their non-grafted counterparts. 

\begin{figure}[h]
\begin{center}
\includegraphics[width=6.5cm,clip]{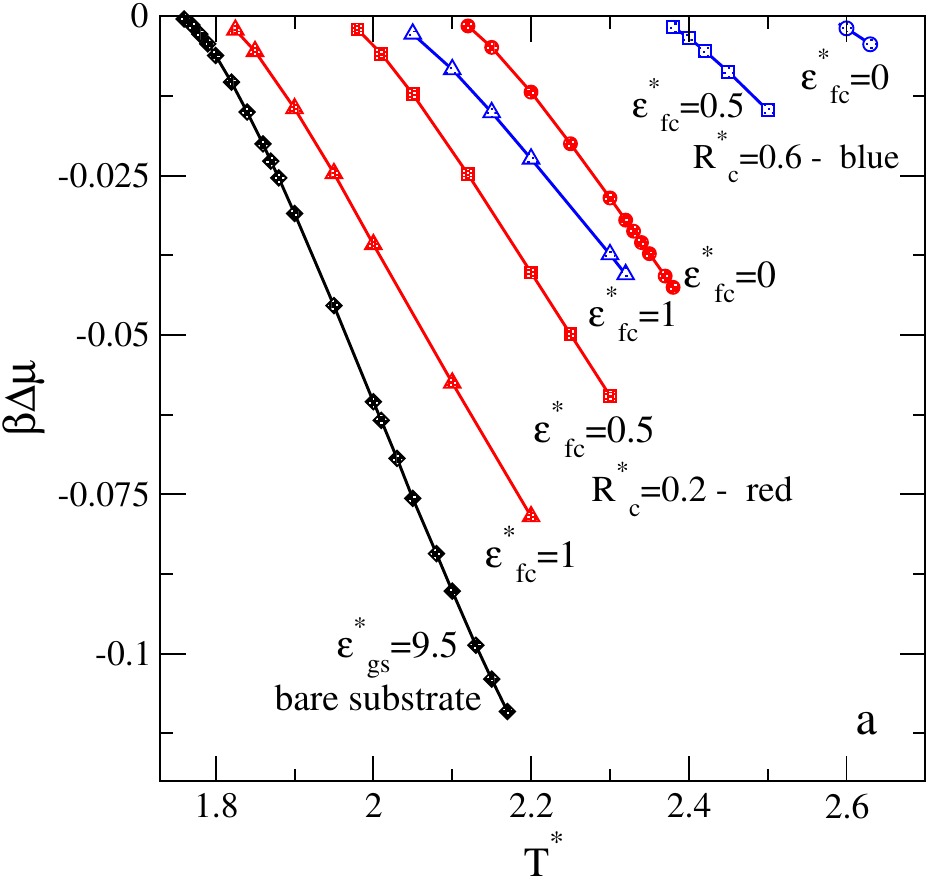}
\includegraphics[width=6.0cm,clip]{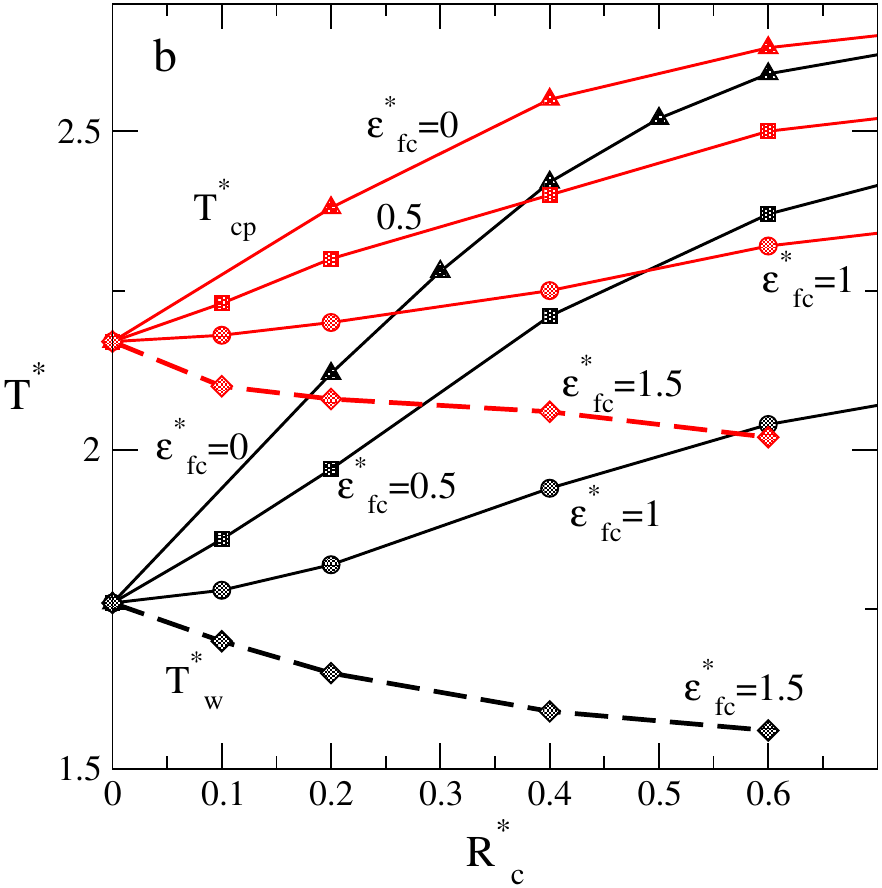}
\end{center}
\caption{(Colour online)
Panel a: Prewetting phase diagrams for water on a solid substrate with grafted monomers
at $R^*_\text{c} =0.2$ (red lines with symbols) and 0.6 (blue lines with symbols).
Circles, squares and triangles correspond to $\varepsilon^*_{\text{fc}}=0, 0.5$   and 1,
respectively. Panel b: Evolution of the wetting temperature, $T^*_\text{w}$,  and of the prewetting critical
temperature, $T^*_{\text{cp}}$, on the grafting density, $R^*_\text{c}$, of monomers.
}
\label{fig2}
\end{figure}

In addition, we observed that the jump of the excess adsorption 
upon the prewetting transition in two types of systems is quite similar even close
to $T^*_{\text{cp}}$, figure~\ref{fig1}c. The corresponding density profiles of water species, before and after the
prewetting transition are shown in figure~\ref{fig1}d. The presence of grafted monomers leads to a
smaller values of the density profile in the region of the first maximum compared to
the bare substrate. This tendency is less pronounced in the vapor phase (prior to the
prewetting transition). By contrast, after the transition, the difference of height
of the profiles is quite big.
Thus, the fluid molecules are ``expelled'' from the vicinity of the substrate surface,
presumably because the grafted monomers play the role of obstacles.
Moreover, if the grafted monomers are present, 
the density profile exhibits a peculiarity at $z \approx 1.5$. 
In summary, augmenting grafting density of monomers, $R^*_\text{c}$,  makes the functionalized substrate
less and less hydrophilic and the equivalency with the decreasing value of $\varepsilon_{\text{gs}}^*$
can be well established. This conclusion is valid for the entire interval of $R^*_\text{c}$,
i.e., up to two-dimensional packing fraction that yields a wetting temperature approaching
the bulk critical temperature.

%%%%%%%%%%%%%%%%%%%%%%%%%%%%%%%%%%%%%%%%%%%%%%
All the results presented in figure~\ref{fig1} were obtained assuming the hard-sphere type
interaction between grafted monomers and water molecules, $\varepsilon_{\text{fc}}^*=0$.
If the grafted monomers attract water species,
the balance of repulsive and attractive forces changes. Consequently, the prewetting phase
diagrams change. It is difficult to choose the values for the parameter $\varepsilon^*_{\text{fc}}$ appropriately
without supporting the experimental evidence to intend reasonable parametrization.
We have chosen $\lambda_{\text{fc}}=1.5\sigma$ fixed, and consider 
the augmenting values for $\varepsilon^*_{\text{fc}}$.
One can intuitively expect that the grafted layers of short molecules may quantitatively
change the trends of behavior of the wetting properties rather than yield qualitative changes.

Two examples illustrating the changes of the prewetting phase diagram, for $R^*_\text{c} =0.2$  and 0.6, are shown
in figure~\ref{fig2}a. 
Apparently, at a value of $\varepsilon^*_{\text{fc}}$ higher than shown for $R^*_\text{c} =0.2$, one can
obtain the wetting temperature even lower, compared to the bare substrate with $\varepsilon_{\text{gs}}^*=9.5$. 
The entire dependence of the wetting temperature, $T^*_\text{w}$,  and of the prewetting critical 
temperature, $T^*_{\text{cp}}$, on the grafting density, $R^*_\text{c}$, with the parameters under study, is
shown in figure~\ref{fig2}b. Both dependencies, $T^*_\text{w}(R^*_\text{c})$ and $T^*_{\text{cp}}(R^*_\text{c})$ are non-linear.
Moreover, at $\varepsilon^*_{\text{fc}} = 1.5$, the wetting temperature decreases upon
increasing $R^*_\text{c}$, indicating that the grafted surface becomes more hydrophilic
than the bare substrate, if water molecules strongly surround or cover the 
attractive monomer obstacles.  
It seems that at this value for $\varepsilon^*_{\text{fc}} = 1.5$,
the system is still above the possible triple point temperature.
One should have in mind, that the water model in question
has been parametrized for temperatures above the triple point temperature. Therefore,
the application of theoretical construction may be questionable at lower temperatures.

\begin{figure}[h]
\begin{center}
\includegraphics[width=6cm,clip]{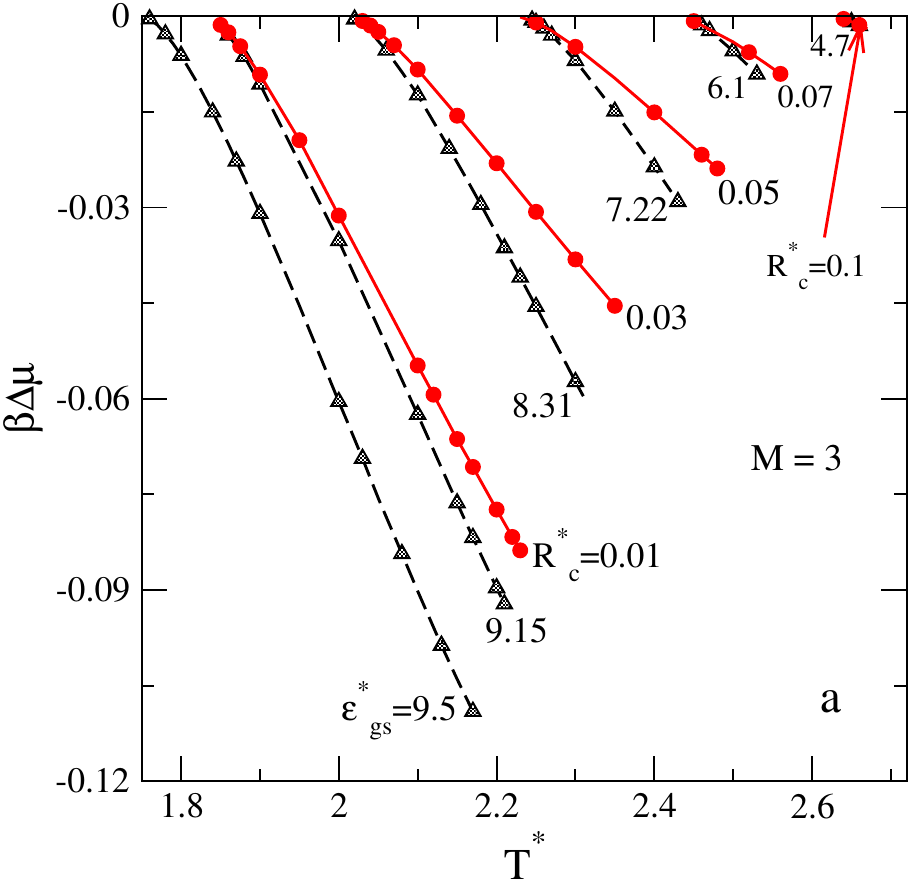}
\includegraphics[width=6cm,clip]{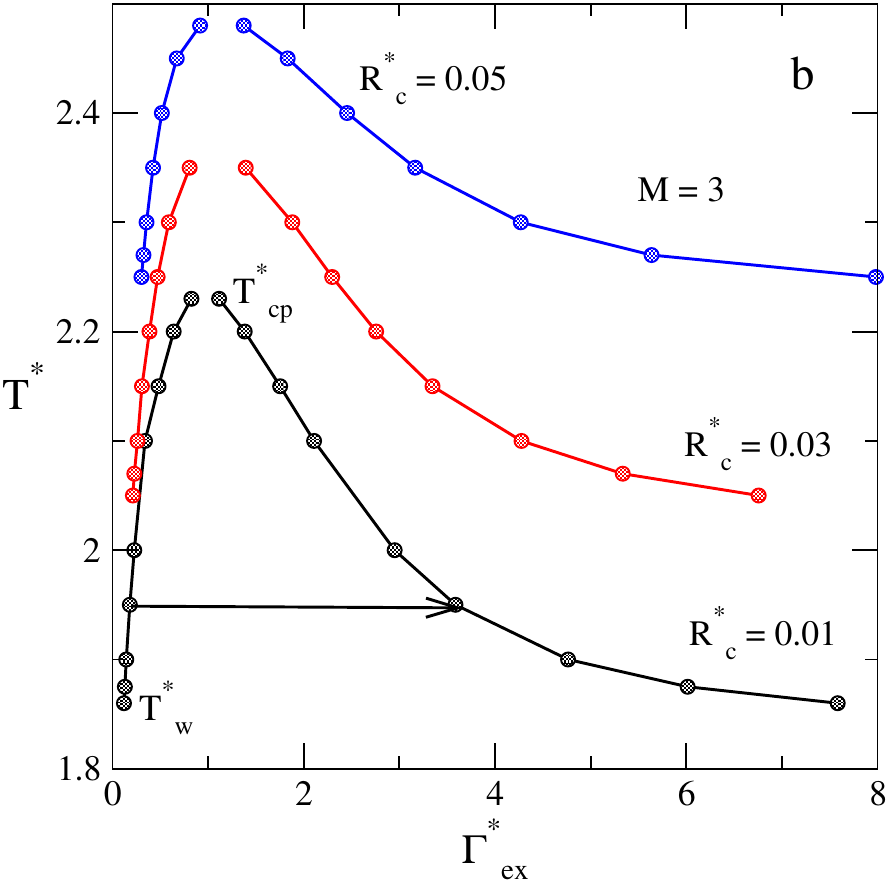}
\includegraphics[width=6cm,clip]{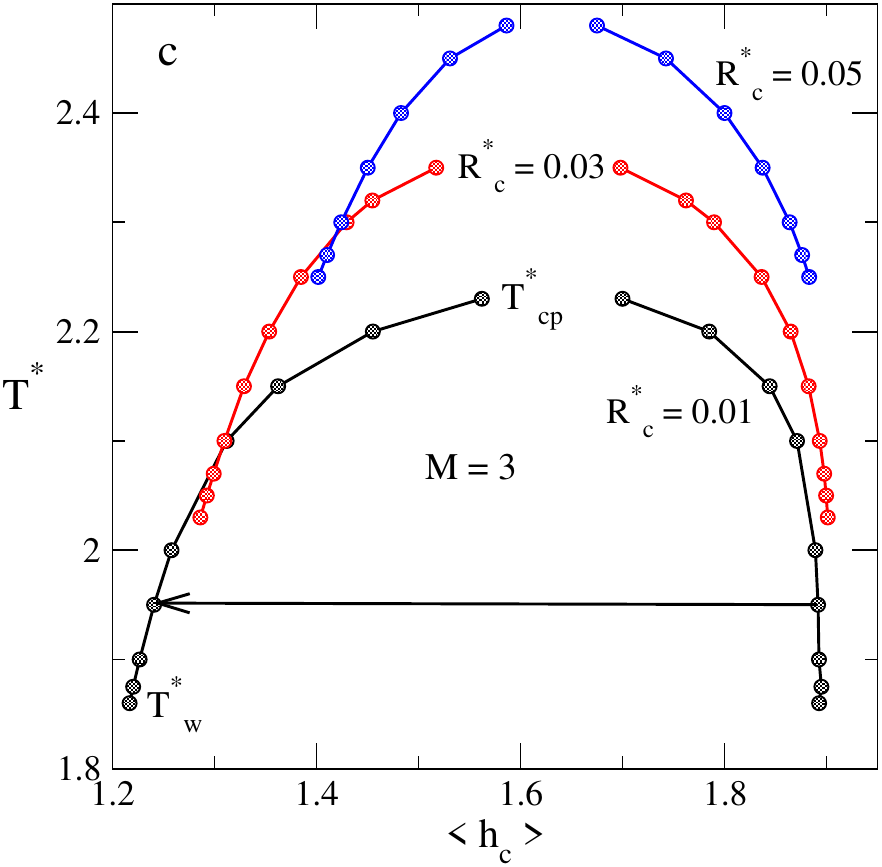}
\includegraphics[width=6cm,clip]{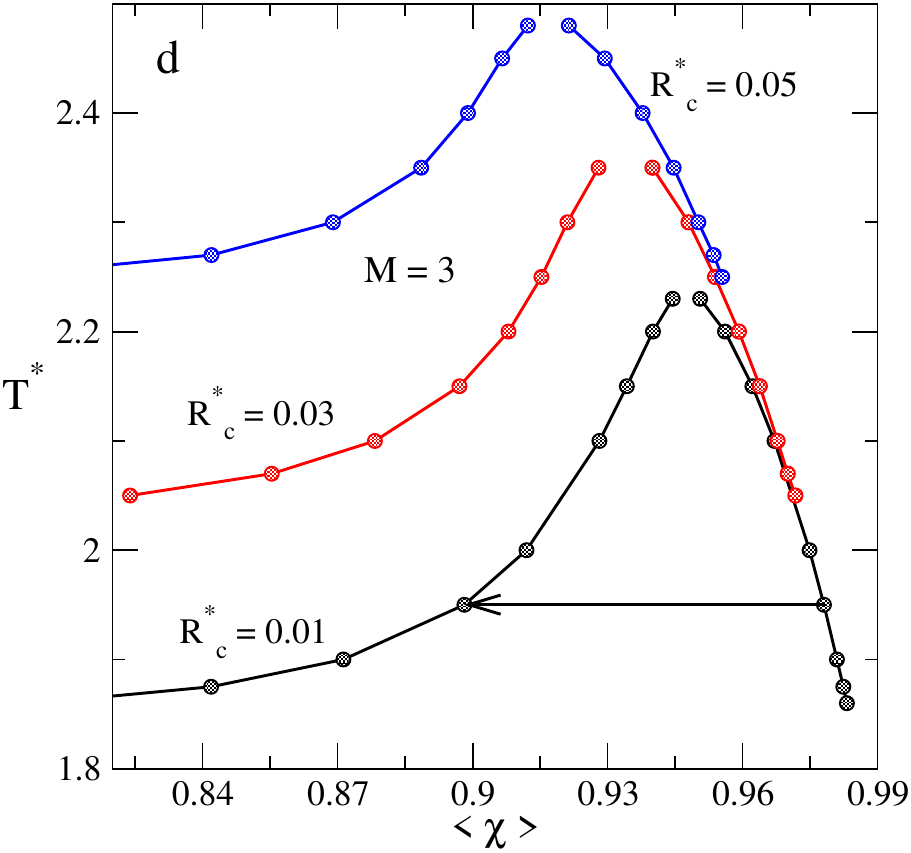}
\end{center}
\caption{(Colour online) 
Panel a:
Prewetting phase diagrams of water on a solid surface with grafted
trimers at a different grafting density, $R^*_\text{c}=$0.01, 0.03, 0.05, 0.07 and 0.1
(from left to right - red lines with circles)
and on ungrafted substrates with a different adsorbing strength,
($\varepsilon_{\text{gs}}^*=9.5$, 9.15, 8.311, 7.22, 6.1 and 4.7) --- black dashed lines 
with triangles. Panels b, c and d:
The $T^* - \Gamma^*_{\text{ex}}$, $T^* - \langle h_c \rangle$ , and $T^* - \langle\chi\rangle$ projections of the
prewetting phase diagrams, respectively.}
\label{fig3}
\end{figure}

A very similar set of calculations were performed for the systems with grafted 
chain particles made of three monomers ($M = 3$). The first series of calculations
were performed for the systems in the absence of segment-water attraction,  $\varepsilon_{\text{fc}}^*=0$.
These short chains were grafted on a strongly adsorbing substrate with $\varepsilon_{\text{gs}}^*=9.5$.
If the grafting density increases from zero  up to $R^*_\text{c} =0.1$, the wetting temperature changes from
its value for a pure substrate, $T^*_\text{w}=1.76$, to  $T^*_\text{w}=2.64$, quite close to the bulk water critical
temperature, figure~\ref{fig3}a. 

Then, we searched for the ungrafted substrates with different $\varepsilon_{\text{gs}}^*$ 
that yield approximately the same $T^*_\text{w}$ for different grafted systems. The entire set of
projection in the chemical potential-temperature plane is shown in figure~\ref{fig3}a. 
Illustration of the magnitude of the jump of excess adsorption upon the
prewetting transition is given in panel b of figure~\ref{fig3}.
%, figure~\ref{fig3}b.
Each of the envelopes starts below from the wetting temperature, $T^*_\text{w}$, and ends up at $T^*_{\text{cp}}$. 
In contrast to the systems with grafted monomers discussed above, now, for $M = 3$, we can construct
the panel describing the conformation changes of grafted species upon the prewetting phase transition.
This is conveniently given in terms of the jump of the first moment of the density profile of chain 
species, $\langle h_c\rangle$, versus reduced temperature, $T^*$, figure~\ref{fig3}c. 
In all cases, we observe that the chains grafted by the first (initial) monomers attain
almost a flat configuration on the solid surface upon the formation of a fluid film, as it follows from
the $\langle h_c\rangle$ values after the prewetting transition, figure~\ref{fig3}c.
Another auxiliary information about the formation of a water film is given in figure~\ref{fig3}d.
The curves witness the changes of the bonding state of molecules upon the prewetting transition.
The branches of the curves to the right of this panel show that the gas-like water vapour contains a
high fraction of non-bonded molecules, high values of $\langle \chi\rangle$. After the transition, a liquid-like film
is formed. The average fraction of non-bonded water molecules is essentially lower in the film, compared to the
vapour phase.

\begin{figure}[h]
\begin{center}
\includegraphics[width=6.5cm,clip]{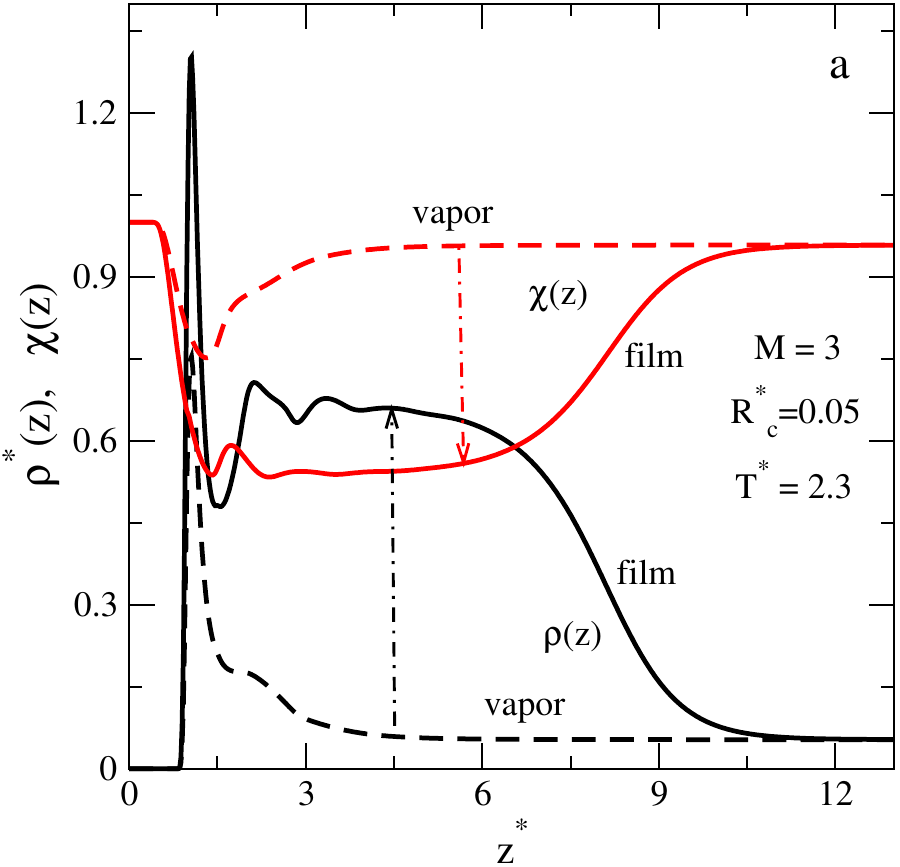}
\includegraphics[width=6.5cm,clip]{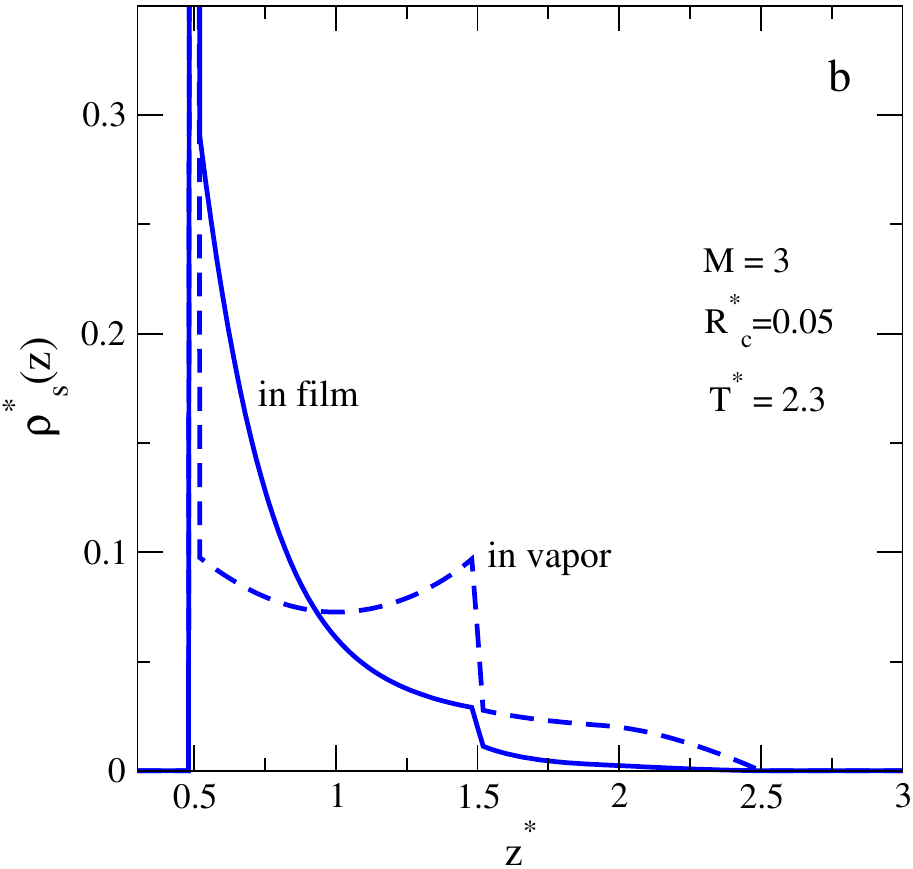}
\end{center}
\caption{(Colour online) Panel a:
The density profiles of water species upon the prewetting transition in the system with
grafted trimers, $M=3$, at $R^*_\text{c} = 0.05$ and $T^* = 2.3$ (black lines).
The profiles of the fraction of non-bonded molecules before and after transition are
shown by red lines.  Panel b: Changes of the profile for grafted molecules upon the
prewetting transition.
}
\label{fig4}
\end{figure}

The phase diagrams in figure~\ref{fig3} can be illustrated in terms of the microscopic structure. One
example of the changes of the density profile of water particles and segments of grafted 
molecules is presented in figure~\ref{fig4}. From figure~\ref{fig4}a, we learn that the
water film after the prewetting transition is rather dense. It involves approximately six layers of 
water molecules whereas the vapor phase before the transition corresponds
to a monolayer with a rather small amount of water particles in the second layer.
The prewetting transition is accompanied by changes of the fraction of non-bonded species.
Vapor phase is mainly composed of non-bonded water molecules. By contrast, the film
after transition contains a large amount of molecules that are bonded between themselves.
Moreover, the formation of a liquid-like film leads to a drastic change of conformation of grafted
trimers. Namely, the segments of trimers are pushed to the surface such that the grafted 
species attain almost a flat configuration after the transition, figure~\ref{fig4}b.

\begin{figure}[h]
\begin{center}
\includegraphics[width=6.5cm,clip]{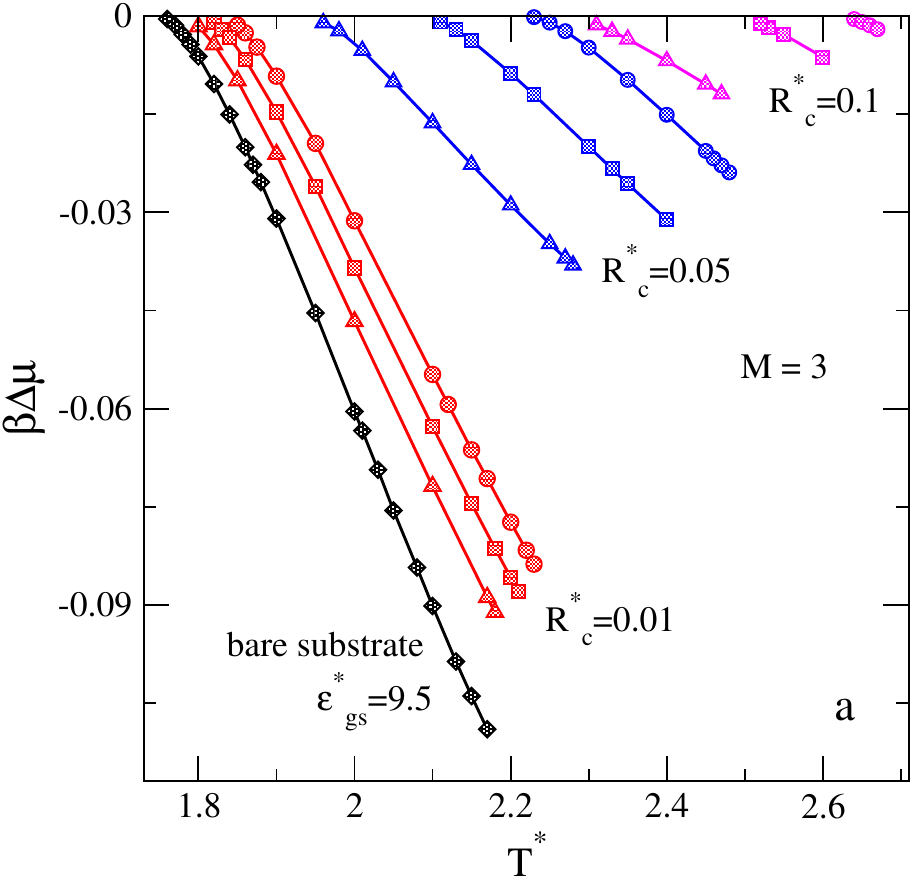}
\includegraphics[width=6.5cm,clip]{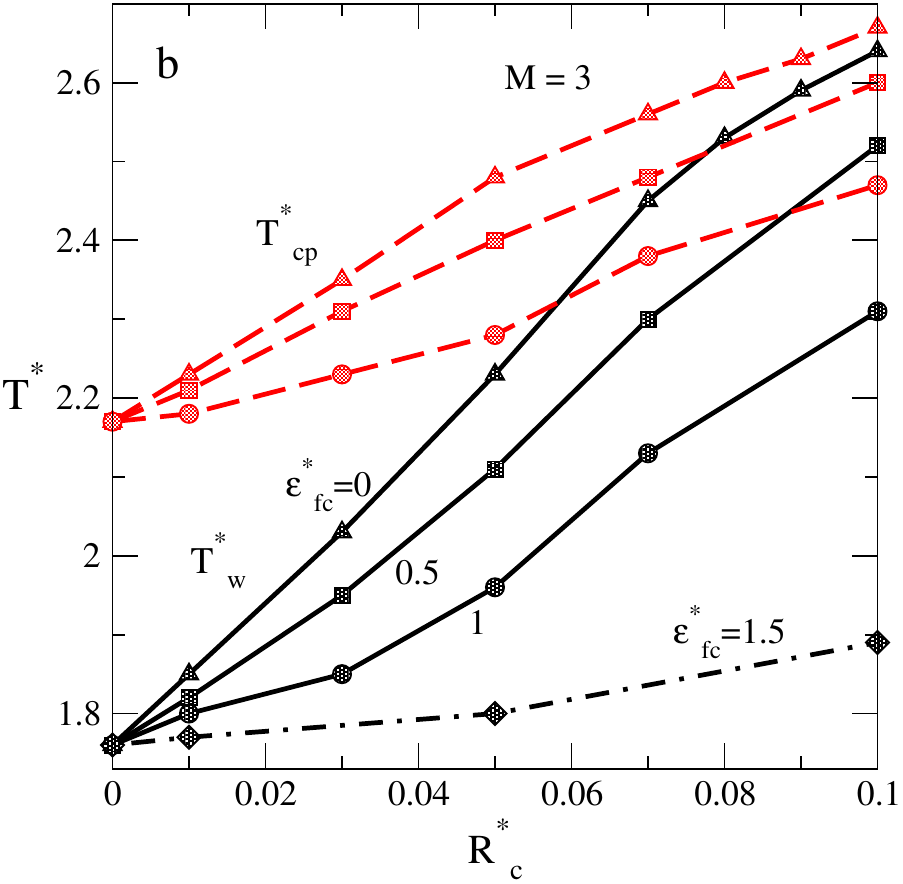}
\end{center}
\caption{(Colour online) 
Panel a: Prewetting phase diagrams of the system with grafted trimers ($M=3$) upon the
changes of the strength of attraction between the grafted species and water molecules,
$\varepsilon^*_{\text{fc}} =0$, 0.5 and 1 (lines with circles, squares and triangles,
respectively. The red, blue and magenta lines correspond to systems at a
different grafting density: $R^*_\text{c} =0.01$, 0.05 and 0.1, respectively.
Panel b: Dependence of the wetting temperature, $T^*_\text{w}$, and of a prewetting
critical temperature, $T^*_{\text{cp}}$, on the grafted density $R^*_\text{c}$.
}
\label{fig5}
\end{figure}

Now, we turn our attention to the effects of attraction between water molecules and 
segments of grafted trimers. Similarly to the model with grafted monomers,
we assume $\lambda_{\text{fc}}=1.5\sigma$, and consider apparently
modest changes $\varepsilon^*_{\text{fc}}$, from 0  to 1.5, as in figure~\ref{fig2}.
A set of results describing the evolution of the prewetting transition lines 
at different values of the grafting density, $R^*_\text{c}$, is shown in figure~\ref{fig5}a.
At a fixed grafting density, the solid surface becomes more attractive
upon increasing $\varepsilon^*_{\text{fc}}$ as expected. However, the magnitude
of the change of the wetting temperature is different and depends on the value
of the grafting density. At higher values of $R^*_\text{c}$, the changes of wettability are
more pronounced. A summarizing insight into the behavior of the wetting temperature and of 
the prewetting critical temperature with the grafting density is provided in figure~\ref{fig5}b.
One can see that the substrates with grafted trimers become more hydrophobic
upon increasing the grafting density  at different values of $\varepsilon^*_{\text{fc}}$
studied. The rate of growth of both characteristic temperatures depends on
the assumed value for the energy $\varepsilon^*_{\text{fc}}$. 
At the highest value under study, $\varepsilon^*_{\text{fc}}=1.5$, the wetting 
temperature still increases, but very slowly. This behavior is in contrast
to the model with grafted monomers, cf. figure~\ref{fig2}b. It seems that in the present case
one needs to consider even higher values for $\varepsilon^*_{\text{fc}}$ to 
observe a decreasing wetting temperature. In other words, a stronger affinity 
between water molecules and segments is needed to cover larger obstacles, $M=3$,
to mitigate their blocking capability for adsorption of water, in comparison to the
model with $M=1$. Entire picture emerging from the consideration of the systems
with grafted trimers and comparison with ungrafted models having approximately
the same wetting temperature shows quantitative differences of adsorption
of water, especially in the interval of temperatures close to the prewetting critical 
temperature.

\begin{figure}[h]
\begin{center}
\includegraphics[width=6.0cm,clip]{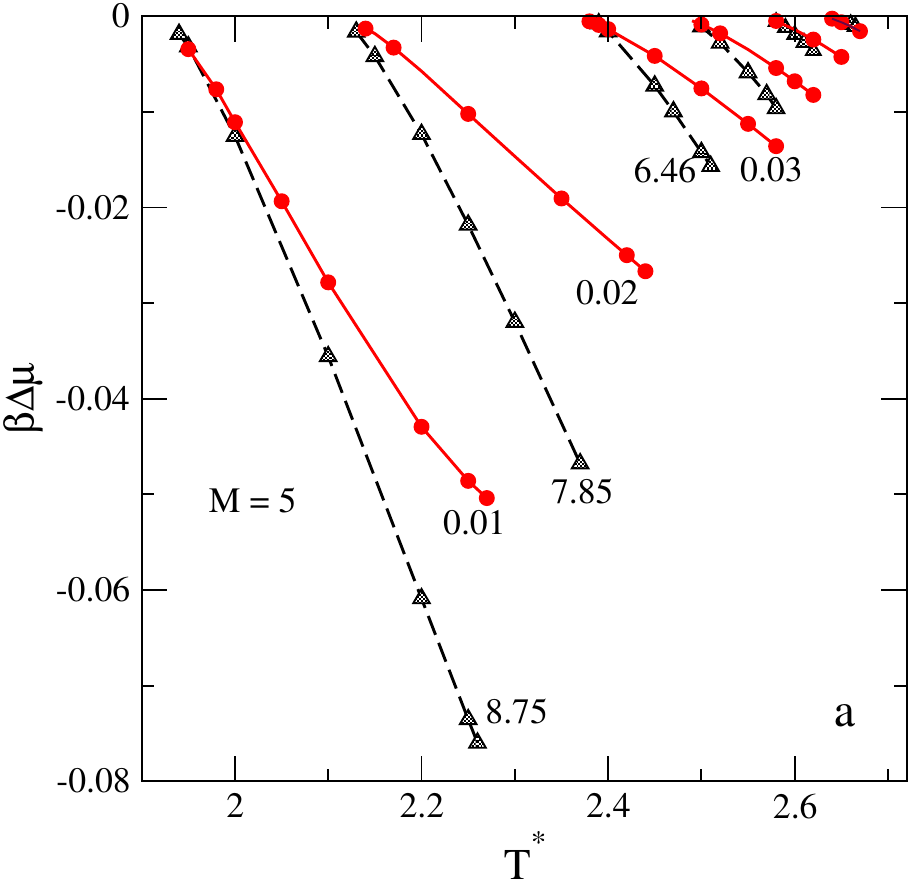}
\includegraphics[width=6.0cm,clip]{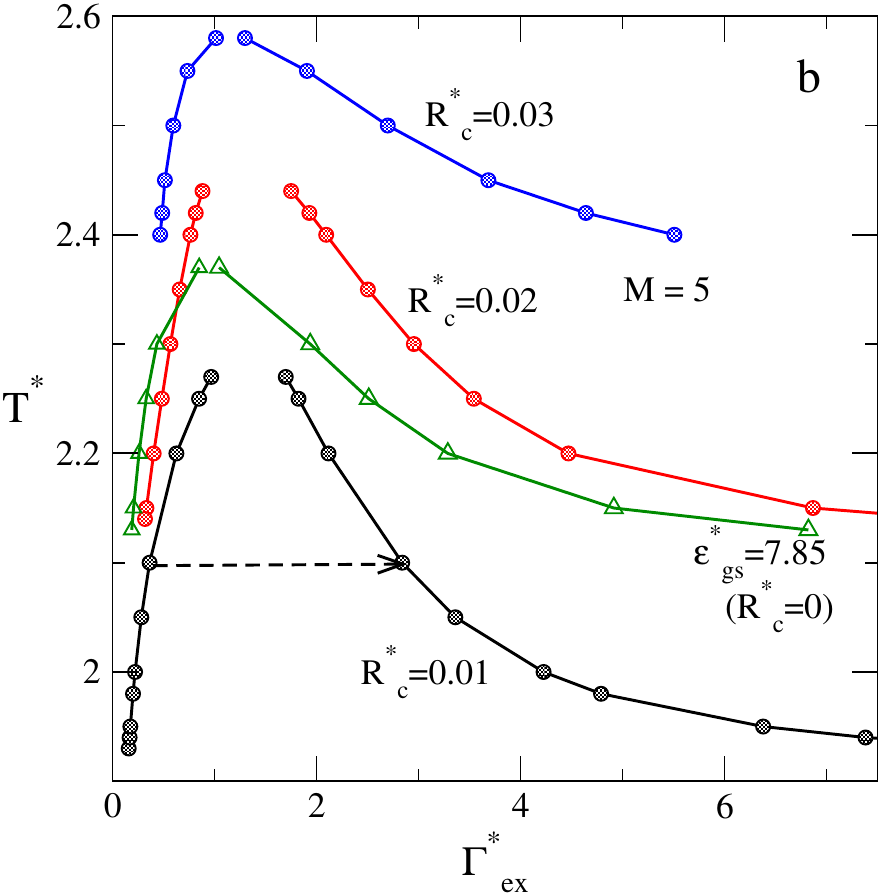}
\includegraphics[width=6.0cm,clip]{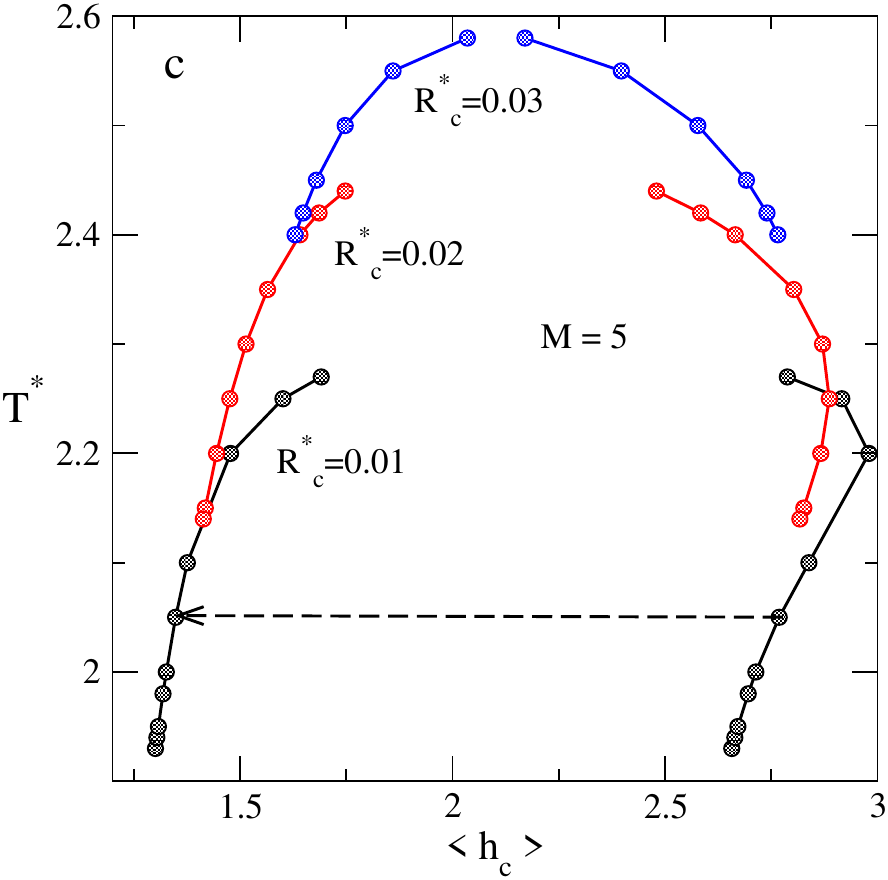}
\includegraphics[width=6.0cm,clip]{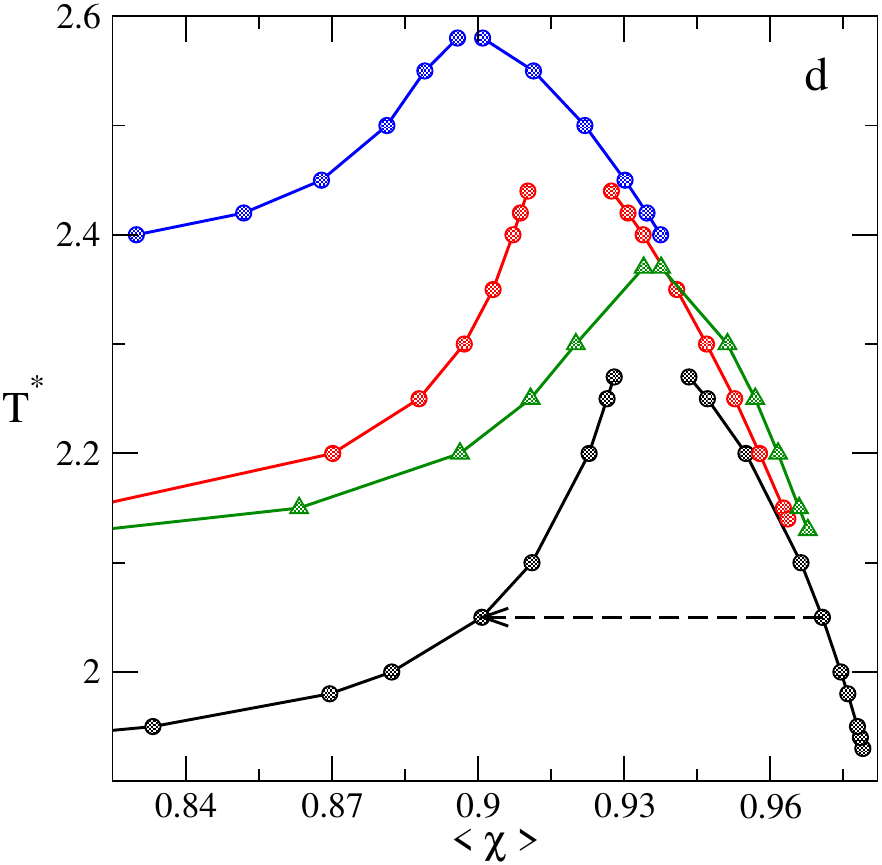}
\end{center}
\caption{(Colour online) Panel a: 
A comparison of the prewetting phase diagrams for the systems with the grafted
pentamers, $M=5$,  at $R^*_\text{c}=0.01$, 0.02, 0.03, 0.035, 0.04 and 0.045 (red
lines with triangles)
and for bare substrates with different adsorbing strength, $\varepsilon^*_{\text{gs}}$,
black lines with circles, at 8.75, 7.85, 6.46, 5.83, 5.23 and 4.6.
Some of these numbers are indicated in the figure for a better identification of the 
curves.
Panels b, c and d:  
The $T^* - \Gamma^*_{\text{ex}}$, $T^* - \langle h_c \rangle$, and $T^* - \langle\chi\rangle$ projections of the
prewetting phase diagrams, respectively. The phase diagram projections for an
ungrafted substrate with $\varepsilon^*_{\text{gs}}=7.85$ are shown in panels b and d for
comparison. The nomenclature of colors is the same in panels b and d.
}
\label{fig6}
\end{figure}

The final stage of our investigation concerns the systems with the grafted pentamers
$M=5$. Again, we begin with the case $\varepsilon_{\text{fc}}^*=0$.
Four panels of figure~\ref{fig6} contain the prewetting phase diagrams of the models
with grafted species (panel a), projections describing the changes of 
excess adsorption (panel b), average height of the grafted layer (panel c) 
and the average fraction of non-bonded species (panel d) upon the prewetting transition.
In addition, we plotted the phase diagrams of the models without grafted 
pentamers at various $\varepsilon^*_{\text{gs}}$, that yield similar values for the wetting temperature,
$T^*_\text{w}$. The highest grafting density at which we identified the wetting temperature
very close to the bulk critical temperature is $R^*_\text{c} = 0.047$. The difference
between the prewetting transition lines of grafted and ungrafted systems that
have a similar wetting temperature is rather big, figure~\ref{fig6}a. Besides, the differences
between the prewetting critical temperatures are well pronounced. 
On the other hand, at a fixed temperature, the jump of adsorption 
upon the prewetting transition in the grafted
system ($R^*_\text{c} = 0.02$) is larger than in the ungrafted system $\varepsilon^*_{\text{gs}}=7.85$
(both have approximately the same wetting temperature). In addition, the magnitude of
change of the average fraction of non-bonded water molecules upon the prewetting transition 
is more pronounced in the grafted system, figure~\ref{fig6}d. It is related to the magnitude of jump
of the excess adsorption during transition. The average height of the grafted layer
changes as well, figure~\ref{fig6}c. If the liquid film is formed,  the grafted layer attains the
configuration in which segments are between mono- and bilayer.

\begin{figure}[h]
\begin{center}
\includegraphics[width=6cm,clip]{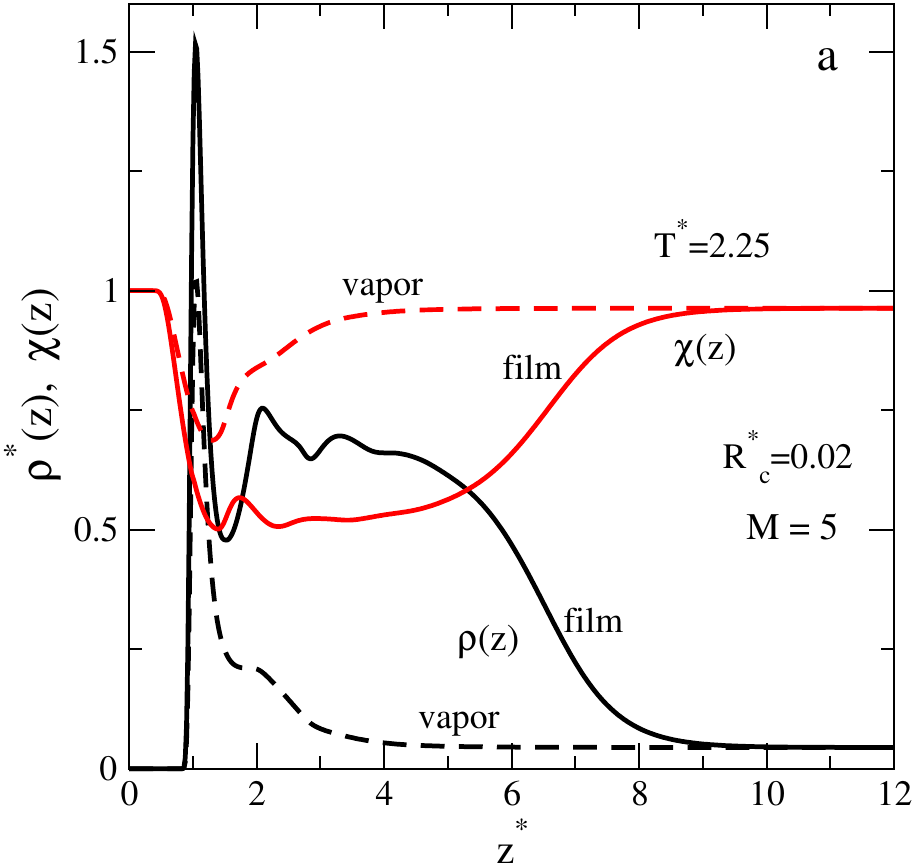}
\includegraphics[width=6cm,clip]{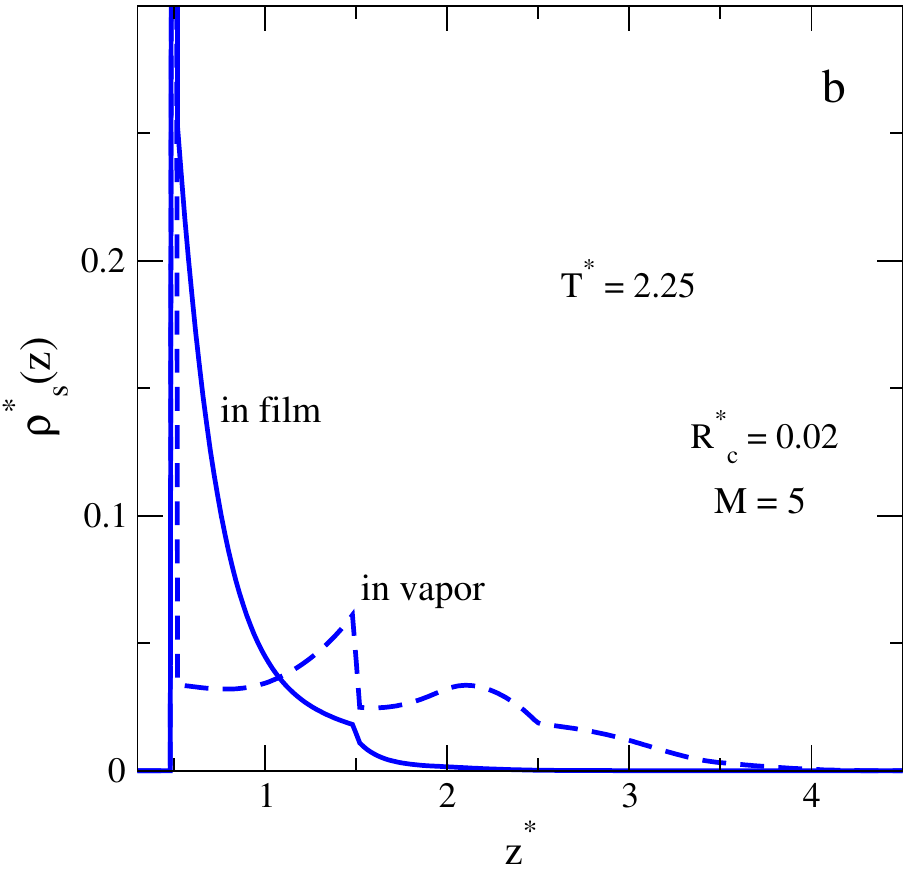}
\end{center}
\caption{(Colour online) Panel a: 
The density profiles of water species upon the prewetting transition in the system with
grafted pentamers, $M=5$, at $R^*_\text{c} = 0.02$ and $T^* = 2.25$ (black lines).
The profiles for the fraction of non-bonded molecules before and after transition are
shown by red lines.  Panel b: Changes of the profile for grafted molecules upon the
prewetting transition.
}
\label{fig7}
\end{figure}

For illustrative purposes, one example of the changes of the density profiles of water
molecules and the total density profile of pentamers upon the prewetting transition
is shown in figure~\ref{fig7}. The profiles for the fraction of non-bonded water molecules
are given in figure~\ref{fig7}a as well. Interestingly, the shrinking of the grafted layer
of pentamers resembles the trends of behavior of grafted trimers, cf. figure~\ref{fig4}b.
In the vapor phase, the profiles are different due to the different number of
segments. However, after the prewetting transition, pentamers and trimers are
pushed into a monolayer type structure, figure~\ref{fig7}b.

\begin{figure}[h!]
\begin{center}
\includegraphics[width=6cm,clip]{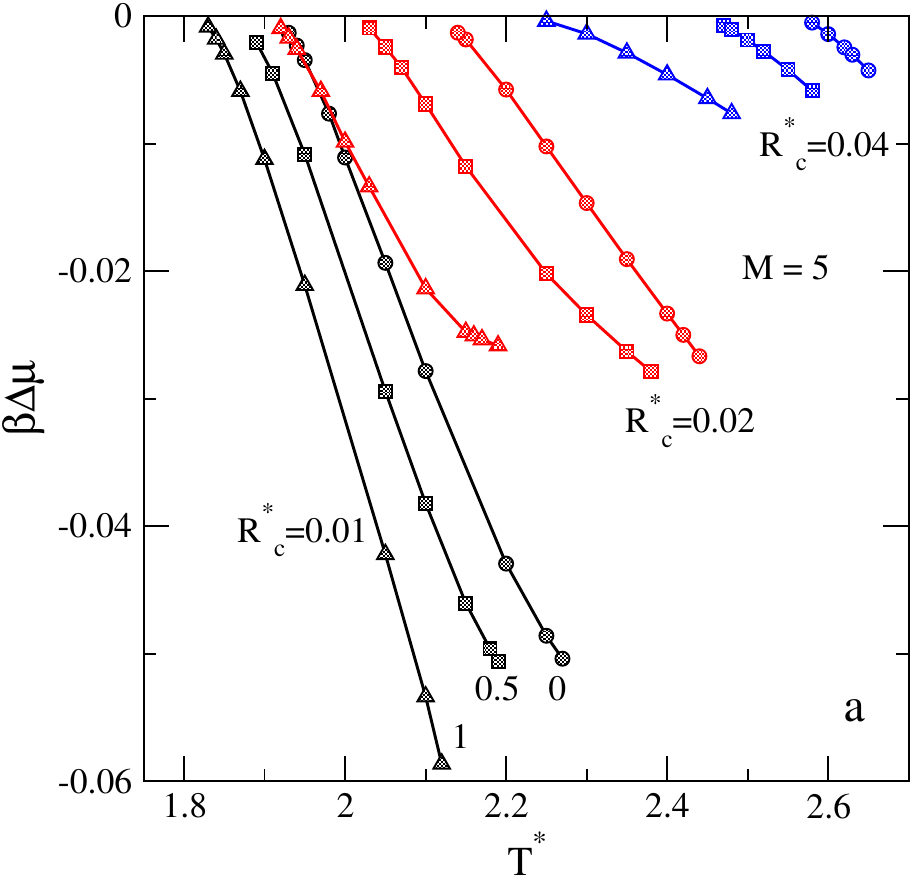}
\includegraphics[width=6cm,clip]{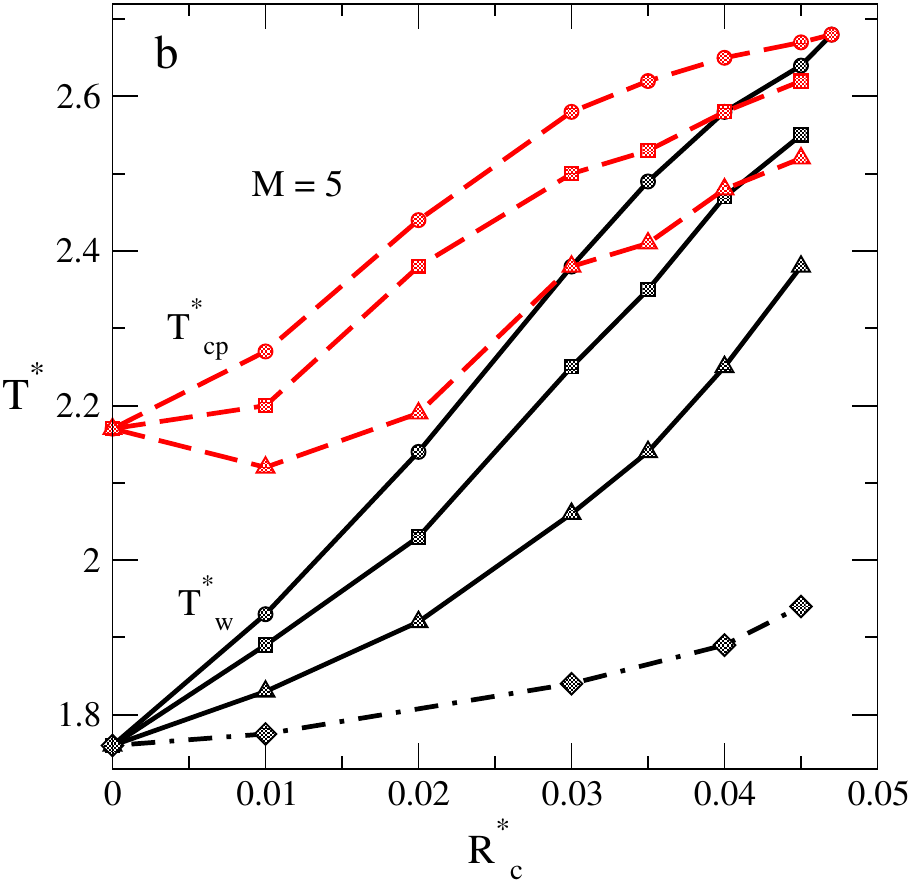}
\end{center}
\caption{(Colour online) 
Panel a: Prewetting phase diagrams of the system with grafted pentamers ($M=5$) upon
changes of the strength of attraction between grafted species and water molecules,
$\varepsilon^*_{\text{fc}} =0, 0.5$ and 1 (lines with circles, squares and triangles,
respectively. The black, red and blue lines correspond to systems at a
different grafting density: $R^*_\text{c} =0.01$, 0.02 and 0.04, respectively.
Panel b: Evolution of the wetting temperature, $T^*_\text{w}$ (black lines), 
and of the prewetting critical temperature, $T^*_{\text{cp}}$ (red lines), 
with grafted density $R^*_\text{c}$. The lines decorated with circles,
squares and triangles correspond to $\varepsilon^*_{\text{fc}}=0$, 0.5 and 1,
respectively. The line with diamonds is for $\varepsilon^*_{\text{fc}}=1.5$.
}
\label{fig8}
\end{figure}

Similarly to our previous discussion, we now turn our attention to the
effects of attraction between water particles and segments of grafted molecules.
The width of the square well attractive potential is $\lambda_{\text{fc}}=1.5\sigma$
as in $M=1$ and $M=3$ systems. However, the energy of interaction, $\varepsilon^*_{\text{fc}}$, changes.
Some examples of the evolution of the prewetting transition lines are given
in figure~\ref{fig8}a. Apparently, the magnitude of changes of $T^*_\text{w}$ and of $T^*_{\text{cp}}$
depends on $R^*_\text{c}$, in non-trivial manner. All the models described in this figure
exhibit an augmenting wettability, if the attractive interaction between segments 
of grafted molecules and water  becomes stronger. Next, we combined the results
for the phase diagrams, at constant $\varepsilon^*_{\text{fc}}$, 
into the plot illustrating the dependence of
characteristic temperatures on the grafting density. These results are
shown in figure~\ref{fig8}b. The conclusions concerning the behavior of the wetting
temperature, $T^*_\text{w}$, in the systems of grafted pentamers are similar
to the trends already observed for grafted trimers, $M=3$, cf. figure~\ref{fig5}b.
If the attraction between the grafted segments and water molecules is strong
enough, e.g., $\varepsilon^*_{\text{fc}}=1.5$, the wetting temperature slowly increases 
upon increasing $R^*_\text{c}$.
If this attraction is weaker, e.g., $\varepsilon^*_{\text{fc}} = 0.5$, the hydrophobicity
of the grafted solid surface grows more rapidly. Apparently, 
for higher values of the parameter $\varepsilon^*_{\text{fc}}$, one may observe
the curve with decreasing values for $T^*_\text{w}$ upon increasing grafted density $R^*_\text{c}$.
At present, we obtained a wavy behavior for the prewetting critical temperature,
$T^*_{\text{cp}}$, with increasing $R^*_\text{c}$ at a moderate attraction strength between 
segments and water molecules (cf. figure~8 of~\cite{dabrowska} for
the behavior of the wetting temperature on the grafting density of monomers from
the evaluation of contact angles).

\section{Summary and conclusions}

We have studied the problem of wetting a solid surface chemically modified
by grafted short chain molecules by water using a simple model
that well captures the bulk phase behavior and applies 10-4-3
interaction potential for water-graphite-type solid interaction.
The chain molecules are modelled as tangentially bonded hard sphere
segments that may interact with water as well.
The theory is a version of the classical density functional approach.
Our principal focus is on the determination of the wetting temperature and
the construction of the prewetting phase diagrams.
The fluid chemical potential-temperature projection of the phase diagram
is evaluated for several grafted systems with monomers, trimers and 
pentamers as the grafted species. 
A detailed comparison is performed with the wetting behavior
of water on differently adsorbing
bare graphite-like surfaces.
It is described how the presence of grafted chains
modify the prewetting transition line.
Other projections for the prewetting transition on temperature,
grafting density, the number of segments and on the attraction strength
between segments of grafted species and water molecules are
obtained and analyzed.
Surface thermodynamic results are illustrated by the fluid density profiles 
and the total segment density profiles in some cases.
All the results are consistent and physically reasonable.
At the moment, we are not able to confront our findings with experimental data.
Probably, the models in question are too simple and need to include various 
elements to make them closer to the laboratory setup.
Nevertheless, the present study can be supplemented by the
calculations of the temperature trends of the contact angle 
to make a link with the data for water on heterogeneous surfaces.
For some systems that involve simple fluids in contact with solids, studied so far, 
predictions of the density functional theory are confirmed by computer simulations.
We expect that similar developments will be performed concerning the phenomena studied
in the present work.

The models of the present investigation permit and require various important
extensions. Undoubtedly, for future studies of selective adsorption one
needs to modify our numerical software to mixtures of water with either organic
co-solvent or solutions of interest in various applications.
Methodologically, it is important to include the effects of chemical 
association between most exposed, terminating segment of grafted 
species with fluid particles. On the other hand, it is of interest to
include more sophisticated interactions between segments belonging to different grafted
molecules. The effects of stiffness of grafted chain species 
are worth taking into account as well. Some of these problems are under study in our laboratory.
These elements, precisely, would contribute to making the present primitive model smarter
and more suitable to find laboratory-type applications.

\section*{Acknowledgements} Author acknowledges helpful discussions with 
Prof. Myroslav Holovko.  Technical support of Magdalena Aguilar at the
Institute of Chemistry of the UNAM is gratefully acknowledged.
Support by CONAHCyT of Mexico under the grant CBF2023-2024-2725 is acknowledged as well.

\ukrainianpart

\title{До побудови мікроскопічної моделі для смарт-покриття твердої поверхні}
\author{О. Пізіо}
\address{Iнститут хiмiї, Нацiональний автономний унiверситет Мексики, Сiркуiто Екстерiор, 04510 Mехіко, Мексика
}

\makeukrtitle

\begin{abstract}
	Підхід функціоналу густини для класичних асоціативних рідин використовується для дослідження фазових діаграм змочування для модельних систем, що складаються з води та графітоподібних твердих поверхонь, хімічно модифікованих невеликою кількістю приєднаних ланцюгових молекул.
	Модель водоподібної рідини запозичено з роботи Кларка та ін. [Mol. Phys., \textbf{104}, 3561 (2006)]. Вона дуже добре описує співіснування води та пари в об'ємі. Кожна ланцюгова молекула складається з тангенціально зв'язаних сегментів твердих сфер. Ми зосереджуємося на дослідженні росту водяної плівки на таких складних підкладках та вивченні поведінки змочування. Для приєднаних мономерів фазові діаграми попереднього змочування подібні до діаграм для води на немодифікованій твердій поверхні. Однак для приєднаних тримерів і пентамерів спостерігається та аналізується набагато цікавіша фізична поведінка. Детально обговорюються тенденції залежностей температури змочування та критичної температури попереднього змочування від густини приєднаних частинок та притягання комплексів води.
	\keywords теорія функціоналу густини, модель води, профілі густини, адсорбція, температура змочування
	
\end{abstract}
\end{document}